\documentclass[useAMS,usenatbib]{mn2e}

\usepackage{graphicx,subfigure}

\usepackage{amssymb,amsmath}

\newcommand{\degree}{\ensuremath{^\circ}}
\title[X-ray bright elliptical galaxies]{Steepening mass profiles,
  dark matter and environment of X-ray bright elliptical galaxies}

\author[P. Das et al.]{Payel Das$^1$\thanks{E-mail:pdas@mpe.mpg.de}, Ortwin Gerhard$^1$, Eugene Churazov$^{2,3}$, Irina Zhuravleva$^2$\\
  $^1$ MPI f\"ur Extraterrestrische Physik, P.O. Box 1603, 85740 Garching, Germany\\
  $^2$ MPI f\"ur Astrophysik, Karl-Schwarzschild-Strasse 1, 85741
  Garching, Germany\\
  $^3$ Space Research Institute (IKI), Profsoyuznaya 84/32, Moscow
  117810, Russia}

\date{Accepted 2010 July 25. Received 2010 July 19; in original form
  2010 April 29}

\begin{document}

\pagerange{\pageref{firstpage}--\pageref{lastpage}} \pubyear{2010}

\maketitle

\label{firstpage}

\begin{abstract}
  We use a new non-parametric Bayesian approach to obtain the most
  probable mass distributions and circular velocity curves along with
  their confidence ranges, given deprojected density and temperature
  profiles of the hot gas surrounding X-ray bright elliptical
  galaxies. For a sample of six X-ray bright ellipticals, we find that
  all circular velocity curves are rising in the outer parts due to a
  combination of a rising temperature profile and a logarithmic
  pressure gradient that increases in magnitude. Therefore at large
  radii, mass density profiles rise more steeply than isothermal
  profiles, implying that we are probing the more massive group-sized
  haloes in which these galaxies are embedded. Comparing the circular
  velocity curves we obtain from X-rays to those obtained from
  dynamical models, we find that the former are often lower in the
  central $\sim 10$ kpc. This is probably due to a combination of: i)
  Non-thermal contributions of up to $\sim 35$\% in the pressure (with
  stronger effects in NGC 4486), ii) multiple-temperature components
  in the hot gas, iii) incomplete kinematic spatial coverage in the
  dynamical models, and iv) mass profiles that are insufficiently
  general in the dynamical modelling. Complementing the total mass
  information from the X-rays with photometry and stellar population
  models to infer the dark matter content, we find evidence for
  massive dark matter haloes with dark matter mass fractions of $\sim
  35$--80\% at $2R_e$, rising to a maximum of 80--90\% at the
  outermost radii. We also find that the six galaxies follow a
  Tully-Fisher relation with slope $\sim 4$ and that their circular
  velocities at $1R_e$ correlate strongly with the velocity dispersion
  of the local environment. As a result, the galaxy luminosity at
  $1R_e$ also correlates with the velocity dispersion of the
  environment. These relations suggest a close link between the
  properties of central X-ray bright elliptical galaxies and their
  environments.
\end{abstract}


\begin{table*}
  \centering
  \begin{tabular}{c c c c c c}
    \hline
    \hline
    Galaxy      &Classification &Distance (Mpc) &$\log L_{\textrm{X}}$ &$\sigma_{\textrm{env}}$ &Source\\
    (1)         &(2)            &(3)            &(4)        &(5)           &(6)\\
    \hline
    NGC 1399 	&cD; E1 pec     & 19.95         &41.63      &370           &D01\\
    NGC 1407 	&E0             & 28.84     	&41.00      &387           &T06\\
    NGC 4472    &E2             & 16.29         &41.43      &607           &G99\\
    NGC 4486  	&E0-E1 pec      & 16.07    	&42.95      &762           &G99\\
    NGC 4649	&E2             & 16.83         &41.28      &702           &G99\\
    NGC 5846 	&E0-E1          & 24.20         &41.65      &322           &M05\\

    \hline
  \end{tabular}
  \caption{Sample of X-ray bright elliptical galaxies: (1) Galaxy name, (2) galaxy classification from NED, (3) distance taken from \protect\cite{tonry+01}, (4) X-ray luminosity taken from \protect\cite{osul+01}, (5) velocity dispersion of the surrounding environment and (6) source of the velocity dispersion values. References are D01 \protect\citep{drink+01}, T06 \protect\citep{tren+06}, G99 \protect\citep{gava+99}, M05 \protect\citep{mahd+05}. \label{tab:sample}}
\end{table*}

\section{INTRODUCTION}\label{sec:intro}

X-ray bright elliptical galaxies are massive galaxies, thought to be
among the most evolved systems in our Universe with a complex
formation history. They are believed to grow from mergers between
smaller galaxies, and after star formation ceases at $z \sim 1$--2, they
become larger and less compact through the significant accretion of
stellar material from neighbouring smaller systems
\citep[e.g.][]{vandokk+08,naab+09}.

Knowledge of the total mass distributions of galaxies gives us insight
into their formation history at several different levels. At the most
basic level, it tells us the combined mass of these collapsed
systems. If we incorporate information from photometry and stellar
population models, we can disentangle the respective luminous and dark
matter components. Comparing their properties with what is expected
from simulations will help place constraints on current theories of
galaxy evolution. The mass profile could also be used as input in
dynamical models of galaxies, therefore mitigating the usual
mass-anisotropy degeneracy. This would provide more stringent
constraints on the orbital structure, which serves as an imprint of
the processes that occurred in the past to create the galaxy. Finally,
it will give us an insight into the relations between the masses of
collapsed systems and the environments they reside in.

There are several methods in the literature for obtaining the mass
distributions of elliptical galaxies. Dynamical models can be
constructed by superposing a library of orbits
\citep[e.g.][]{rix+97,geb+03,thomas+04,cap+06} or distribution
functions \citep[e.g.][]{dejong+96,gerhard+98,kron+00}, or by
constructing a system of particles \citep[][]{lorenzi+08,lorenzi+09}
such that the projection of the system best reproduces observed
surface-brightness and kinematic profiles. These models give the mass
distribution and orbital structure of the galaxies
simultaneously. Strong lensing gives the projected mass within the
Einstein ring and weak lensing studies provide mass density profiles
for a sample of galaxies
\citep[e.g.][]{treu+04,mand+06,gava+07,koop+09}. The properties of
X-ray bright elliptical galaxies allow an additional possibility for
obtaining mass distributions
\citep[e.g.][]{nuls+95,fuk+06,chur+08,nag+09}. The X-ray spectra are
dominated by lines and by continuous emission from thermal
bremsstrahlung radiation produced in the surrounding halo of hot
gas. The spectra can be deprojected and fitted to derive 3-D
temperature and density profiles of the gas, and if we assume it is in
hydrostatic equilibrium, we can derive the enclosed mass profile.

Generally the literature points towards isothermal mass distributions
in elliptical galaxies, with a conspiracy between the luminous and
dark matter components resulting in flat circular velocity curves
similar to that found in spiral galaxies \citep[e.g.][]{kron+00,
  koop+09, chur+10}. \cite{fuk+06} however, determined mass
distributions of a sample of 53 elliptical galaxies from Chandra and
XMM-Newton observations, and found that they are better described by
power laws with an index $\gtrsim 1$, i.e. their circular velocity
curves range between flat and rising. The particle-based dynamical
models of \cite{lorenzi+08,lorenzi+09} found falling circular velocity
curves for the intermediate-luminosity galaxies, NGC 3379 and NGC
4697. \cite{thomas+07} found a range of slopes in the elliptical
galaxies belonging to the Coma cluster, from their orbit-based
dynamical models.

In this paper, we examine the isothermality of mass distributions from
X-ray observations and whether they are accurate enough to derive dark
matter mass profiles, use in dynamical modelling, and determine global
properties of elliptical galaxies. To address these issues we need to
apply hydrostatic equilibrium using a procedure that is as free from
systematic biases as possible.

\cite{cowie+87,hump+06,fuk+06,nag+09} used methods where the measured
temperature and density profiles are parameterised before applying
hydrostatic equilibrium, to circumvent differentiating the observed
profiles, which are often noisy. \cite{chur+08} avoided
differentiation by interpolating to obtain only the potential
profile. \cite{nuls+95} found the most likely mass profile of NGC 4486
(M87) from temperature and density profiles using a non-parametric
method, and \cite{hump+09} used a parametric Bayesian analysis. The
parametric methods used in the literature will generally underestimate
the range of mass profiles consistent with the data. They also
introduce systematic biases in the masses derived because of the
assumptions on the profile shapes.

We use the sample of galaxies from \cite{chur+10} and we discuss them
and the properties of the hot X-ray gas they harbour briefly in
Section \ref{sec:sample}. In Section \ref{sec:mass} we describe the
implementation of a new non-parametric Bayesian approach to obtain the
total mass and circular velocity profiles from hydrostatic
equilibrium. In Section \ref{sec:tests} we show the results of tests
of the new method and how to optimise it. In Section \ref{sec:app} we
show the total mass profiles and circular velocity curves we obtain by
applying the method to the sample of galaxies, and the stellar and
dark matter contributions we infer from published photometry and
stellar population model mass-to-light ratios. In Section
\ref{sec:discuss} we compare the individual circular velocity curves
we obtain to previous determinations from X-rays and published
dynamical models. We also examine the isothermality of the mass
profiles of these galaxies, and look for any correlations that may
exist between the stellar component, the total circular velocities and
properties of the surrounding environment. We end with our conclusions
in Section \ref{sec:conc}.

\section{THE SAMPLE }\label{sec:sample}

We work on the sample of galaxies analysed in \cite{chur+10}. Table
\ref{tab:sample} lists the galaxies in the sample, the galaxy
classification, the distances we assume, the X-ray luminosity and the
velocity dispersion of the surrounding environment. NGC 1399 is
located at the centre of the Fornax cluster, NGC 1407 and NGC 5846 at
the centre of groups, and NGC 4472, NGC 4486 and NGC 4649 at the
centre of sub-clumps in the irregular Virgo cluster. NGC 4486 is at
the centre of the most massive sub-component. \cite{drink+01}
determined the velocity dispersion of the environment surrounding NGC
1399 from 92 galaxies within a radius of about 2.7$\degree$ (930 Mpc),
\cite{tren+06} from 35 galaxies within about 900 kpc of NGC 1407 and
\cite{mahd+05} from 87 galaxies within a radius of 1.8$\degree$ (777
kpc) of NGC 5846. \cite{gava+99} calculated the velocity dispersion in
sub-clumps containing NGC 4472, NGC 4486 and NGC 4649 using 166, 62
and 58 galaxies respectively.

The sample galaxies harbour significant amounts of hot gas that except
in the case of NGC 4486, appears relatively undisturbed, suggesting
hydrostatic equilibrium.

\begin{figure*}
\centering
\includegraphics[width=18cm]{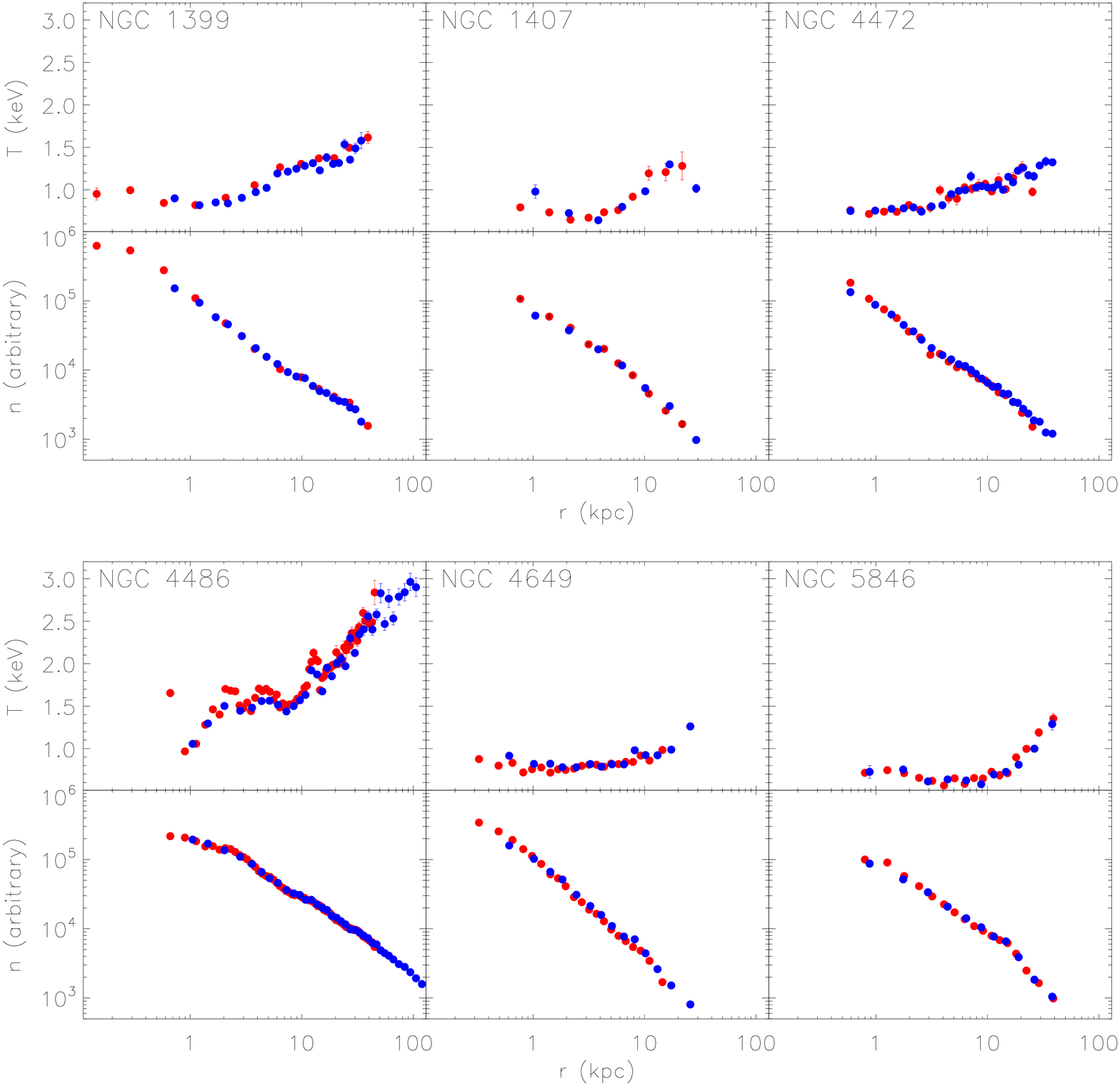}
\caption{Properties of the hot gas surrounding the sample of six X-ray
  bright elliptical galaxies: Deprojected temperature (upper plot) and
  density profiles (lower plot) were obtained by
  \protect\cite{chur+10} from the analysis of Chandra observations
  (red dots) and XMM-Newton observations (blue dots). The temperature
  is shown on a linear scale. The density and radii are shown on
  logarithmic scales. \label{fig:xray}}
\end{figure*}

\subsection{Density and temperature profiles of the hot
  gas}\label{sub:data}

The X-ray spectrum of the hot gas surrounding X-ray bright ellipticals
consists of many emission lines and continuous emission primarily from
the mechanism of thermal bremsstrahlung. To obtain information on the
temperature and density profiles of the gas, the spectra are fitted
with models that make assumptions about the absorption along the
line-of-sight and the metal abundance in the gas.

We use the deprojected temperature and density profiles obtained by
\cite{chur+10} from Chandra and XMM-Newton observations, shown in
Figure \ref{fig:xray}. The composite profiles benefit in the central
region from the high spatial resolution of Chandra and in the outer
parts from the large field-of-view and collecting power of
XMM-Newton. The XMM-Newton profiles are not more spatially extended
than the Chandra profiles as one would normally expect, because
several pointings were used with the Chandra instrument.

Churazov et al. first spherically deprojected the X-ray spectra using
a non-parametric least-squares algorithm described in
\cite{chur+03}. This finds the set of 3-D spectra in spherical shells
that project to best-fit the observed spectra, assuming a power-law
decline in the emission outside the maximum observed radius. The
deprojected spectra were then fit with the single-temperature APEC
code in XSPEC, which resulted in a determination of the 3-D
temperature and density for each shell. The abundance of heavy
elements in the model was fixed at 0.5 solar in all shells. This was
assumed because for cool gas with approximately solar abundance of
heavy elements, the contributions of continuum and lines are difficult
to disentangle unambiguously, resulting in an anti-correlation between
the emission measure and abundance.

Figure \ref{fig:xray} shows generally a good agreement between the
Chandra and XMM-Newton profiles in the region of overlap. NGC 1399,
NGC 1407, NGC 4649 and NGC 5846 have deprojected temperature profiles
that start around 1.0 keV in the central region, dip slightly towards
0.5 keV and then slowly increase outwards to about 1.5 keV. The
temperature profile of NGC 4472 increases steadily from about 0.7 keV
to 1.4 keV while the temperature profile of NGC 4486 increases much
more steeply from about 1 keV to 3 keV with a small dip in
between. The density profiles are much smoother than the temperature
profiles. They appear to be very similar to each other with a linear
decrease in a log-log scale and are therefore close to power laws.

\section{NON-PARAMETRIC RECONSTRUCTION OF THE MASS
  DISTRIBUTION}\label{sec:mass}

If the gas is in hydrostatic equilibrium, we can obtain the mass
distribution from the deprojected gas temperature and density
profiles. Here we describe a new non-parametric Bayesian approach to
obtain the most probable mass profiles within some confidence range.

\subsection{Hydrostatic equilibrium}\label{sub:nonparam}

Hydrostatic equilibrium relates the 3-D temperature and density
profiles of the gas to the 3-D mass distribution of the galaxy. If the
gas is relatively undisturbed, we can assume it is in hydrostatic
equilibrium. Balancing the pressure of the gas with the gravitational
forces acting on it in a spherical system gives:

\begin{equation}\label{eq:he}
  -\frac{d\Phi(r)}{dr} = \frac{1}{\rho(r)}\frac{dP(r)}{dr}
\end{equation}

\noindent where $\rho$ is the gas mass density, $P$ the gas pressure
and $r$ is the 3-D radius from the centre of the galaxy. We assume
that the gas is ideal and therefore $P = nk_BT$ and $\rho = \mu m_p
n$, where $n$ is the particle number density of the gas, and $\mu =
0.61$ is the average gas particle mass in terms of the proton mass,
$m_p$. This value of $\mu$ corresponds to a helium number density of
$7.92\times 10^{-2}$ and 0.5 solar abundance of heavier elements. Now
we have related the temperature and density of the gas to the
gravitational potential it resides in. We can express Equation
\eqref{eq:he} in terms of the circular velocity curve, a
distant-independent measure of the mass, $V_c^2 = r d\Phi/dr$. The
circular velocity at some radius in a galaxy is the orbital velocity
of a star at that radius on a circular orbit in the same gravitational
field. Equation \eqref{eq:he} becomes:

\begin{equation}\label{eq:vc}
  V_c^2 = - \frac{k_b T}{\mu m_p} \frac{d\ln P}{d\ln r}
\end{equation}

\noindent It can be seen that to obtain the circular velocity curve,
the temperature and gradient of the logarithmic pressure is
required. This is more convenient than the usual form of this
equation, where both the temperature and density derivatives are
required.

\subsection{Bayesian approach}\label{sub:bayes}

We would like to find the most probable logarithmic pressure gradients
(and then circular velocity curve from hydrostatic equilibrium) within
some confidence range given the deprojected temperature and density
profiles measured from the data.  \cite{merritt+94} showed that it is
best to treat these problems non-parametrically because parametric
methods are susceptible to systematic biases and underestimate the
confidence ranges. The procedure we develop can eventually incorporate
the spectral fitting and deprojection directly, but in a first step we
concentrate here on generalising parametric methods on deprojected
profiles to non-parametric ones. We adopt a Bayesian approach rather
than the more conventional method of $\chi^2$ minimisation for two
reasons: Firstly, prior information such as the intrinsic smoothness
of the profiles is easily incorporated. Secondly, the probabilistic
nature of Bayesian methods means confidence ranges are more easily
extracted.

Let us describe the galaxy by some model $M$. Bayes' theorem tells us
the probability of $M$ given the deprojected temperature and density
profiles $X$ from the X-ray observations:

\begin{equation}
  p(M \mid X) = \frac{p(X \mid M) \, p(M)}{p(X)}
\end{equation}

\noindent Taking the logarithm of both sides gives:

\begin{equation}
  \ln(p(M \mid X)) = \ln(p(X \mid M)) + \ln(p(M)) - \ln(p(X))
\end{equation}

\noindent $p(M \mid X)$ is the posterior probability of the model
given the deprojected temperature and density profiles. $p(X \mid M)$
is the likelihood probability $\mathcal{L}(M)$ of the deprojected
temperature and density profiles given the model. $p(M)$ is the prior
probability of the model, indicating what we thought about the
probability of that model before we knew anything about the
deprojected temperature and density profiles. Finally $p(X)$ is the
probability of those particular deprojected density and temperature
profiles, but here only acts as a normalising factor because the
profiles are fixed by the X-ray observations. To obtain the most
probable model profile, we need to maximise the posterior probability
$p(M \mid X)$. In order to find all model profiles within some
confidence range we have to find the shape of the posterior
probability distribution.

We define our model $M = M(T, lpg, P, V_c, m)$, where $T$ is the
temperature, $lpg$ is the logarithmic pressure gradient, $P$ is the
pressure, $V_c$ is the circular velocity and $m$ is the enclosed
mass. The pressure $P$ is obtained by integrating the logarithmic
pressure gradient $lpg$. The circular velocity $V_c$ is obtained by
applying hydrostatic equilibrium, and the mass $m$ follows from $V_c^2
= Gm/r$. The model $T$ and $P$ are compared with the deprojected
$T_{\textrm{depro}}$ and $P_{\textrm{depro}}$ obtained from the X-ray
observations and deprojection. The model functions are defined on a
grid of $n_{\textrm{mod}}$ logarithmically-spaced radii $r$, finer
than the grid of $n_{\textrm{depro}}$ deprojected radii
$r_{\textrm{depro}}$. We interpolate the functions to the grid of
deprojected radii for the calculation of the $\chi^2$ function
below. We shall use $j$ to denote an element on the model grid and $k$
to denote an element on the deprojected grid. For example, $T[j]$
denotes the model temperature at the radius $r[j]$ on the model grid
and $T[k]$ denotes the model temperature at the radius
$r_{\textrm{depro}}[k]$ on the deprojected grid.

We assume that the likelihood function $\mathcal{L}(M)$ takes the form
of a multi-variate Gaussian:

\begin{equation}	
  \mathcal{L}(M) \propto \exp(-\chi^2/2)
\end{equation}
 
\noindent $\chi^2$ is the goodness-of-fit of the model temperature to
the deprojected temperature and the model pressure to the deprojected
pressure:

\begin{equation}
  \chi^2 = \sum_{k=1}^{n_{\textrm{depro}}} \left(\frac{P_{\textrm{depro}}[k] - P[k]}{\epsilon_P[k]}\right)^2 + \sum_{k=1}^{n_{\textrm{depro}}} \left(\frac{T_{\textrm{depro}}[k] - T[k]}{\epsilon_T[k]}\right)^2
\end{equation}

\noindent where $\epsilon_P$ and $\epsilon_T$ are 1-$\sigma$
statistical errors on the deprojected pressure and temperature
profiles respectively. This choice for the likelihood function assumes
a Gaussian error distribution and can be generalised if needed.  To
enforce physically acceptable solutions, we impose the following
boundary conditions as priors on the model temperature and logarithmic
pressure gradient profiles:

\begin{enumerate}
\item{$T \ge 0$}
\item{$V_c \in \Re \Rightarrow lpg \le 0$} 
\end{enumerate}

We also need to define a smoothing prior because unlike in a
parameterised model, the best-fit non-parametric model would go
exactly through all the deprojected temperature and density points
resulting in noisy profiles. Also, as the model grid is finer than the
deprojected grid, without smoothing there would be many solutions
where the model is very similar to the deprojected temperature and
density at the deprojected radii but with varying values in
between. Therefore applying smoothing constraints reduces the model
degeneracies by getting rid of unphysical solutions. We specifically
penalise non-smooth solutions of the model temperature, logarithmic
pressure gradient and resulting circular velocity curve because this
is consistent within the hydrostatic framework. We define the
smoothing prior as follows:

\begin{equation}
  p(M) \propto \exp(-S)
\end{equation}

\noindent $S$ is a weighted sum of the mean-square second derivatives
of the temperature, logarithmic pressure gradient and circular
velocity curve:

\begin{align}
	&S = c( S_1 + S_2 + S_3)\\
	&S_1 =  \sum_{j=2}^{n_{\textrm{mod}}-1}\left(\frac{\ln T[j+1]-2
	\ln T[j] + \ln T[j-1]}{h^2}\right)^2\\
    	&S_2 =  \sum_{j=2}^{n_{\textrm{mod}}-1}\left(\frac{lpg[j+1]-2
        	lpg[j] + lpg[j-1]}{h^2}\right)^2\\
    	&S_3 =  \sum_{j=2}^{n_{\textrm{mod}}-1}\left(\frac{\ln V_c[j+1]-2
        	\ln V_c[j] + \ln V_c[j-1]}{h^2}\right)^2
\end{align}

\noindent $h$ is the logarithmic interval between the model radii in
an equally-spaced logarithmic radial grid and $c =
n_{\textrm{depro}}\lambda/(n_{\textrm{mod}} - 2)$ determines the
weight of the smoothing term relative to the $\chi^2$ term. It is
defined so that for fixed $\lambda$, the smoothing term is independent
of $n_{\textrm{depro}}$ and $n_{\textrm{mod}}$. $\lambda$ is specified by the user and
Section \ref{sub:smooth} describes how we choose its optimal value. As
the second derivatives are difficult to estimate at the boundaries, we
omit their contribution to the smoothing prior.

Combining the likelihood function and the priors we get the following
expression for the logarithm of the posterior probability, which we
call $F$:

\begin{equation}\label{eq:F}
  F = \ln(p(M \mid X)) = -\frac{\chi^2}{2} - S + N
\end{equation}

\noindent $N$ is a normalisation constant, which we set to zero for
convenience.

\subsection{The distribution of posterior probabilities}\label{sec:dist}

Now that we have defined our model and the posterior probability of a
particular model given the X-ray information, we need an initial model
and a procedure to generate new models in order to find the
distribution of posterior probabilities. Here we will discuss our
initial model and the `proposal function', a function used to generate
a new model. We then describe an algorithm we implement, optimised to
find the most probable model. This is similar to the method of
\cite{mag99}, who found the most probable 3-D distribution of stars
given a projected surface-brightness distribution. We also describe a
second algorithm based on the Metropolis-Hastings sampling scheme in a
Markov Chain Monte Carlo (MCMC) approach, optimised to extract
confidence ranges associated with a model. We define an `iteration' as
a change proposed by the proposal function that lies within the
boundary conditions.

\subsubsection{The initial model}\label{subsub:initial}

The initial model is given by $M_{\textrm{init}}(T_{\textrm{init}},
lpg_{\textrm{init}}, P_{\textrm{init}}, V_{\textrm{c,init}},
m_{\textrm{init}})$, calculated on the model grid. $T_{\textrm{init}}$
is determined by fitting a highly smoothed spline to the deprojected
temperature $T_{\textrm{depro}}$. $lpg_{\textrm{init}}$ is set to the
slope of a straight-line fit between the logarithm of the deprojected
pressure, $P_{\textrm{depro}}$ and the logarithm of the deprojected
radii, $r_{\textrm{depro}}$. $V_{\textrm{c,init}}$ and
$m_{\textrm{init}}$ follow from applying hydrostatic equilibrium
(Equation \eqref{eq:vc}), and the posterior probability of this
initial model $F_{\textrm{init}}$ can be calculated using Equation
\eqref{eq:F}.

\subsubsection{The proposal function}\label{subsub:prop_func}

The proposal function $q(M_{i+1} \mid M_i)$ is a probability
distribution that generates a model $M_{i+1}$ from model $M_i$. We
only make changes to the model $T$ and $lpg$ and then calculate the
resulting model $P$ from integrating $lpg$ and the model $V_c$ and $m$
by applying hydrostatic equilibrium. The proposal function is defined
as:

\begin{equation}\label{eq:prop_func}
  q(M_{i+1} \mid M_i) = p[j] \, G_T[j] \, G_{lpg}[j] 
\end{equation}

\noindent $p[j] = 1/n_{\textrm{mod}}$, is the probability of picking
point $j$ on the model grid. $G_T[j] = p(T_{i+1}[j] \mid T_i[j])$ is
the probability of going from $T_i[j]$ to $T_{i+1}[j]$ and $G_{lpg}[j]
= p(lpg_{i+1}[j] \mid lpg_i[j])$ is the probability of going from
$lpg_i[j]$ to $lpg_{i+1}[j]$, at the chosen grid point $j$. $G_T[j]$
and $G_{lpg}[j]$ are described by Gaussians centred on $T_i[j]$ and
$lpg_i[j]$ respectively and with dispersions $\sigma_T[j]$ and
$\sigma_{lpg}[j]$. $\sigma_T$ is initially set to the root-mean
squared deviation between $T_{\textrm{init}}$ and
$T_{\textrm{depro}}$, and $\sigma_{lpg}$ is initially set to the
root-mean squared deviation between $lpg_{\textrm{init}}$ and a
two-point estimate of the logarithmic pressure gradient.

\begin{figure*}
  \centering
  \includegraphics[width=18cm]{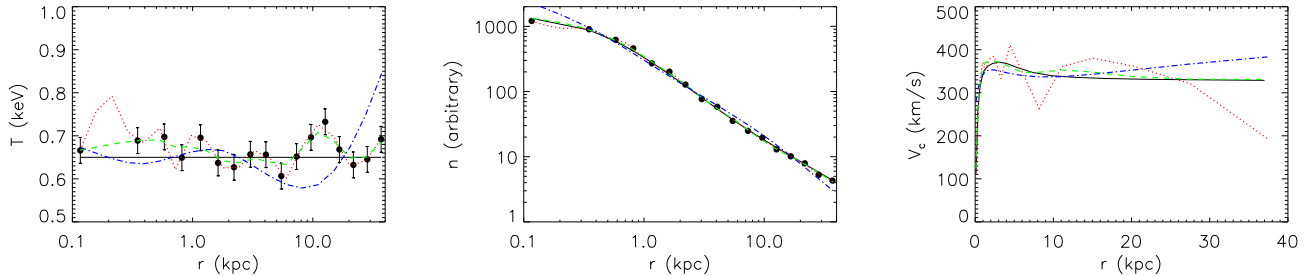}
  \caption{Application of the non-parametric analysis to a test model:
    The black lines show the (i) temperature, (ii) density and (iii)
    circular velocity curves for our test model. The filled black
    circles show the pseudo temperature and density profiles. The
    remaining lines show recovered profiles from pseudo deprojected
    temperature and density profiles generated from the test model,
    assuming $\lambda = 0.001$ (red, dotted), $\lambda = 0.5$ (green,
    dashed) and $\lambda = 100$ (blue,
    dash-dotted).\label{fig:test_model}}
\end{figure*}

\subsubsection{Obtaining the most probable solution (non-Markov
  mode)}\label{subsub:most_prob}

To obtain the most probable model we want to get to the maximum of the
posterior probability distribution as quickly as possible. We first
relax the initial model to mitigate any bias introduced in choosing
it. In this phase the proposal function is used to make a change to
the model at a random point on the model grid. If the boundary
conditions are met, then the proposed change is accepted and the new
posterior probability $F$ and change in posterior probability $\Delta
F$ are calculated. After $n_{\textrm{relax}}$ iterations, the
relaxation phase ends and the average change in the posterior
probability $\langle \Delta F \rangle$ is calculated. In the
relaxation phase, it is possible for $\langle \Delta F \rangle$ to be
either negative or positive as we accept all changes subject to the
boundary conditions.

In the improvement phase, a change to the model is proposed via the
proposal function and if the boundary conditions are met then an
acceptance probability $\alpha$ is specified for going from model
$M_i$ to model $M_{i+1}$:

\begin{equation}\label{eq:acc_prob}
  \alpha(i,i+1) = \min(1,r) \textrm{ where } r = \exp{\frac{\Delta F}{\langle \Delta F \rangle}}
\end{equation}

\noindent A random number $z$ is generated between 0 and 1. If $z \le
\alpha(i,i+1) $, the proposed change is accepted and the dispersion of
the proposal function is decreased, $\sigma_T[j] = \sigma_T[j]/u$ and
$\sigma_{lpg}[j] = \sigma_{lpg}[j]/u$, where $u=1.01$ to allow for
fine changes. Otherwise the proposed change is rejected and
$\sigma_T[j] = \sigma_T[j] \times u$ and $\sigma_{lpg}[j] =
\sigma_{lpg}[j] \times u$. This update to the proposal function is
made after each change to the model so that in general we follow a
path that maximises the posterior probability, but still allow for the
possibility of jumping away from a local maximum. The value for
$\langle \Delta F \rangle$ is updated every $n_{\textrm{log}}$ steps.

This procedure is continued until the model no longer significantly
changes.

\subsubsection{Obtaining the confidence range (Markov
  mode)}\label{subsub:conf_range}

To obtain the confidence range we need to probe the shape of the
posterior probability distribution. The most efficient way to do this
is using the Metropolis-Hastings algorithm, which generates models
with a probability equal to the posterior probability. In our
implementation of this algorithm we have three phases. The first is a
relaxation phase of the initial model as in Section
\ref{subsub:most_prob} with $n_{\textrm{relax}}$ iterations. The
second is a tuning of the proposal function, where the proposal
function is updated as in the second phase in Section
\ref{subsub:most_prob}. This phase ends after $n_{\textrm{tune}}$
steps, defined such that the acceptance rate in the third phase is
~23\%. This is recommended as the most efficient rate for probing the
shape of the posterior probability distributions in high-dimensional
implementations of the Metropolis-Hastings algorithm \citep{lid09}.

In the third phase we want to map out the shape of the posterior
probability distribution around the maximum. The proposed changes must
therefore satisfy `detailed balance':

\begin{equation}\label{eq:detail_bal}
  p(M_i \mid X) \, p(M_{i+1} \mid M_i) = p(M_{i+1} \mid X) \, p(M_i \mid M_{i+1})
\end{equation}

\noindent $p(M_{i+1} \mid M_i) = q(M_{i+1} \mid M_i) \, \alpha(i,i+1)$
and $p(M_i \mid M_{i+1}) = q(M_i \mid M_{i+1}) \, \alpha(i+1,i)$,
where the proposal function $q$ and acceptance probability $\alpha$
were defined in Equations \eqref{eq:prop_func} and \eqref{eq:acc_prob}
respectively. This is ensured by no longer changing the proposal
function and replacing the definition of $r$ from Equation
\eqref{eq:acc_prob} with:

\begin{equation}
 	r = \exp{\Delta F}
\end{equation}

\noindent $r$ is called the Metropolis ratio. This will produce a
sequence of models called a Markov chain. This is continued until the
Markov chain converges to the posterior probability distribution. The
initial $\sim 25$--50\% of the accepted changes is called the `burn-in
phase' and is discarded because they do not reflect the posterior
probability distribution. The convergence of the Markov chain can be
checked by looking at the distribution of model values at each grid
point and ensuring that subsequent iterations produce insignificant
changes.

Marginalisation with a Markov chain is trivial because the density of
any grid point in the Markov chain will be proportional to the
posterior distribution marginalised over the other variables. The
shape of the marginal posterior density at each grid point will tell
us the median and confidence ranges, and the expectation value is
simply a mean of all the values generated in the Markov chain.

\section{TESTS}\label{sec:tests}

We need to test the ability of the two methods to recover the circular
velocity curve of a galaxy and to choose values for the parameters
$n_{\textrm{relax}}$, $n_{\textrm{log}}$, $n_{\textrm{burnin}}$,
$n_{\textrm{iter}}$ and $\lambda$ that we have introduced in Section
\ref{sub:bayes}, \ref{subsub:most_prob} and \ref{subsub:conf_range}.

For the relaxation phase we set $n_{\textrm{relax}} =
3(2n_{\textrm{mod}})$, similar to that used by \cite{mag99}. In the
non-Markov mode we set $n_{\textrm{log}} = 8(2n_{\textrm{mod}})$,
again similar to that used by \cite{mag99}. In the Markov mode we set
$n_{\textrm{burnin}}$ to 25\% of the total number of accepted changes
to ensure we are only sampling the posterior probability
distribution. We choose the total number of accepted changes
$n_{\textrm{iter}}$ as the number of iterations after which the model
no longer appears to change significantly.

To test the two modes of operation and calibrate the number of
accepted changes $n_{\textrm{iter}}$ and the smoothing parameter
$\lambda$, we define a model of a typical X-ray bright elliptical
galaxy, from which we draw pseudo 3-D temperature and density
profiles.

\begin{figure}
\centering
\includegraphics[width=8.5cm]{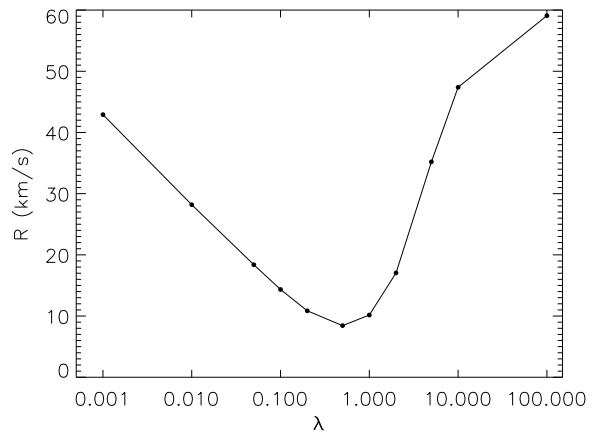}
\caption{Residuals between the circular velocity curve of the test
  model and the circular velocity curves recovered by the
  non-parametric analysis for different values of the smoothing
  parameter, $\lambda$. \label{fig:smooth}}
\end{figure}

\subsection{The test model and pseudo temperature and density
  profiles}

The test model is defined by a 3-D temperature profile, 3-D density
profile and a circular velocity curve $[T_{\textrm{test}},
n_{\textrm{test}}, V_{\textrm{c,test}}]$ typical for a X-ray bright
elliptical galaxy. These profiles are shown by the black lines in
Figure \ref{fig:test_model}. For the circular velocity curve we use
the dynamical model of NGC 5846 by \cite{kron+00}. We assume the
temperature $T_{\textrm{test}}$ is constant at 0.65 keV and then
calculate the corresponding $lpg_{\textrm{test}}$ and density
($n_{\textrm{test}}$) profiles. $n_{\textrm{test}}^2$ is projected and
pure Poissonian statistics typical for XMM-Newton is added. The
projected density profiles are then deprojected as in \cite{chur+08}
and a deprojected density profile is obtained. We added Gaussian
random deviates to the temperature assuming a constant error of 0.03
keV to complete the set of pseudo deprojected profiles shown with
filled black circles in Figure \ref{fig:test_model}.

\begin{figure*}
\centering
\subfigure[]{
\includegraphics[width=8.5cm]{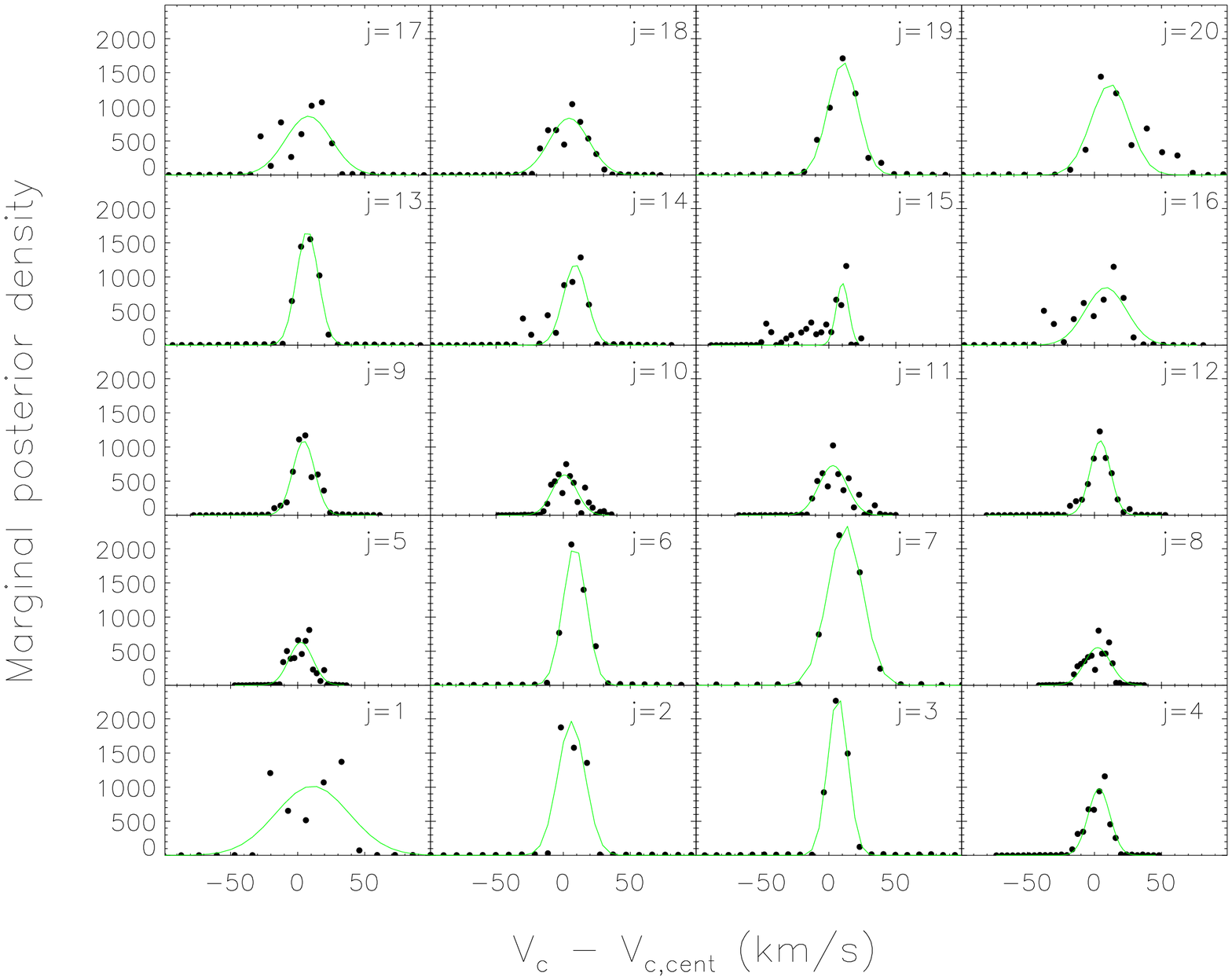}
	}
\subfigure[]{
	\includegraphics[width=8.5cm]{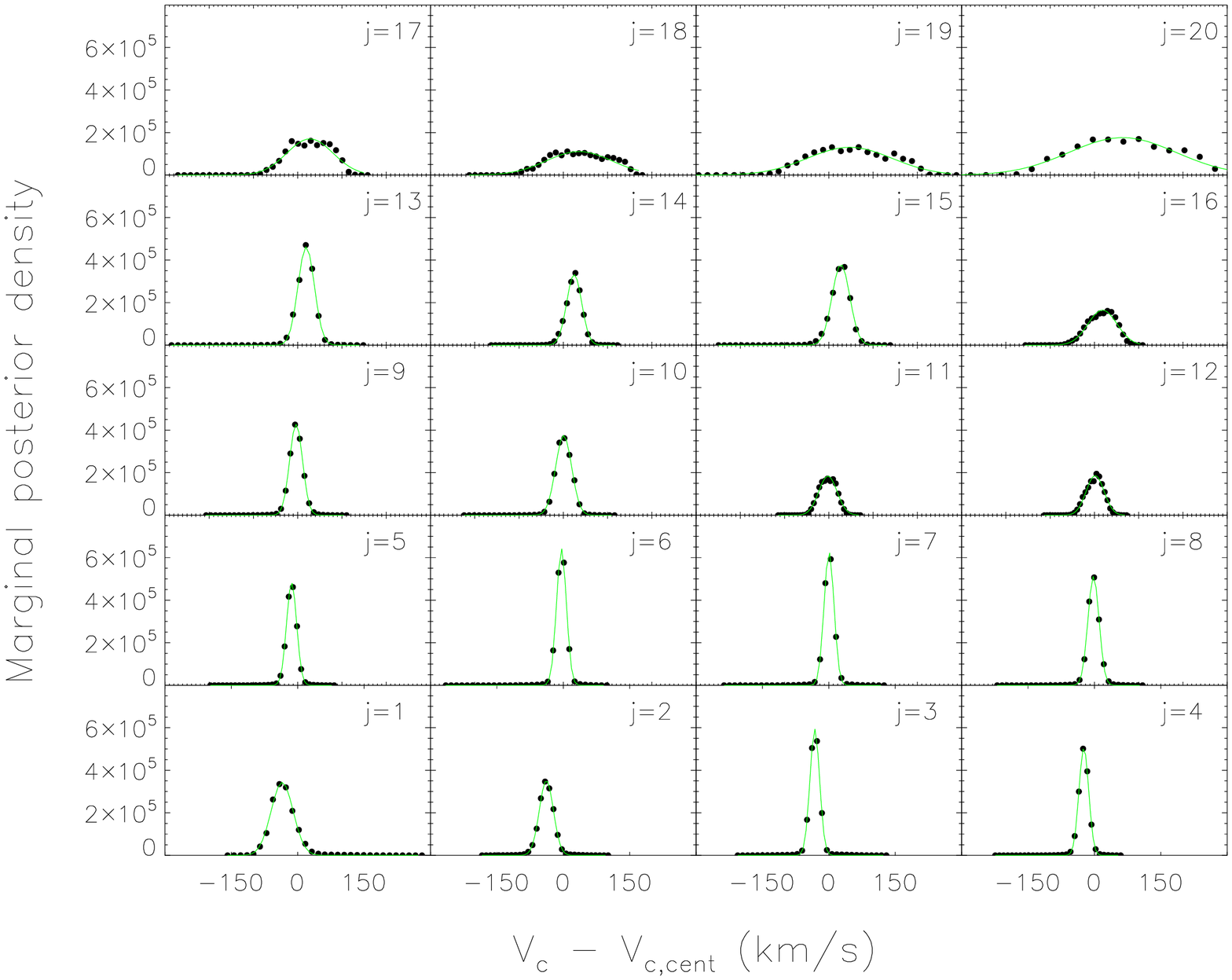}
	}
        \caption{The circular velocity values $V_c$ generated at each
          grid point $j$ from our non-parametric analysis of the
          pseudo deprojected temperature and density profiles: The
          filled black circles show the binned number density
          (proportional to the marginal posterior density) of $V_c$
          values at each grid point $j$ for (a) $n_{\textrm{iter}} =
          10^4$ and (b) $n_{\textrm{iter}} = 2 \times 10^6$ and the
          green lines show the best-fit
          Gaussians.\label{fig:marg_post}}
\end{figure*}

\begin{figure*}
\centering
\includegraphics[width=18.0cm]{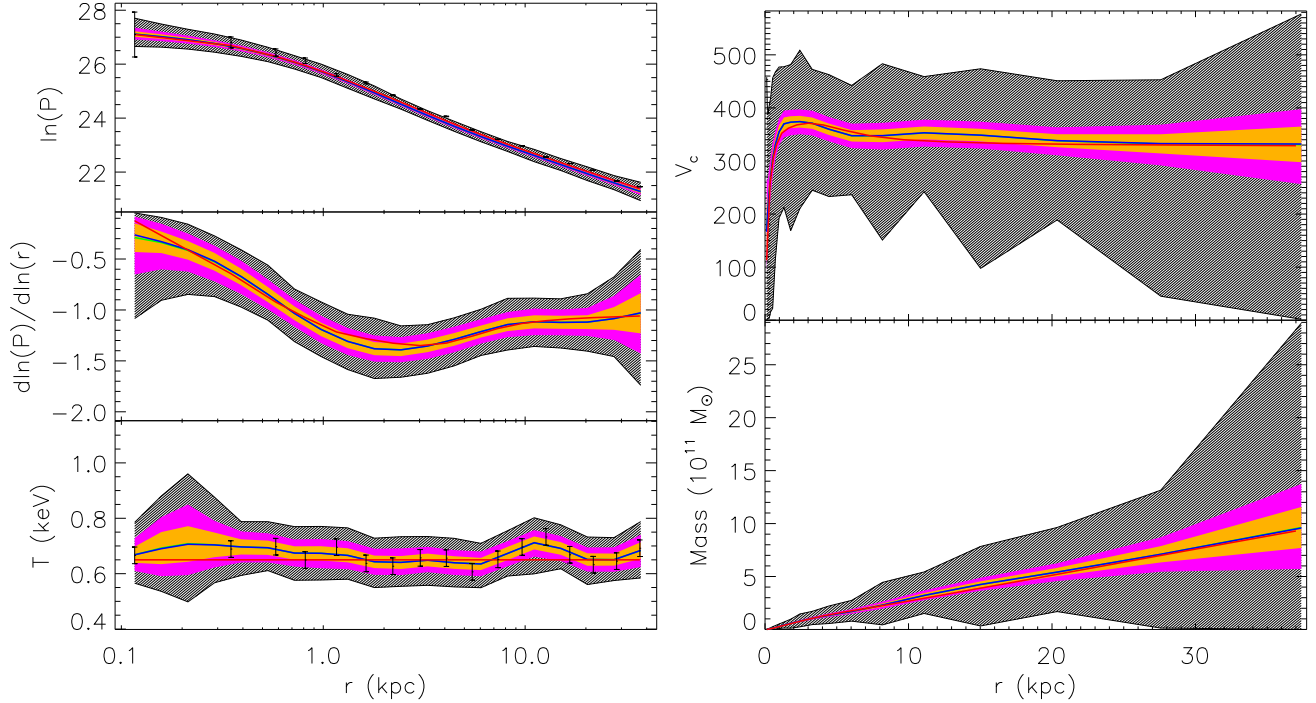}
\caption{Application of the non-parametric analysis to the pseudo
  deprojected temperature and density profiles: Going
  counter-clockwise from the top left are the (i) logarithmic
  pressure, (ii) logarithmic pressure gradient, (iii) temperature,
  (iv) mass and (v) circular velocity profiles. The left three
  profiles are on a logarithmic radial scale, while the two profiles
  on the right are on a linear radial scale. The red lines correspond
  to the input test model, the green lines correspond to the recovered
  expected profiles and the blue lines correspond to the recovered
  median profiles. The grey region shows all generated points, the
  pink region shows the 95\% confidence range and the orange region
  shows the 68\% confidence range. \label{fig:bayes}}
\end{figure*}

\begin{table*}
  \centering
  \begin{tabular}{c c c c c c c c}
    \hline
    \hline
    Galaxy    &$\chi_T^2$ &$\chi_{lnP}^2$ &$M_{25} \, (10^{11}M_{\odot})$ &$V_{c,25}$ (km/s)     &$\zeta$           &$\zeta_{<10}$    &$\zeta_{>10}$    \\
    (1)       &(2)       &(3)            &(4)                      &(5)                  &(6)               &(7)              &(8)             \\
    \hline
    NGC 1399  &2.7        &10.1          &$13.2 \pm 1.1$           &$475.0 \pm 20.9$     &$1.15 \pm 0.03$   &$1.16 \pm 0.06$  &$1.60 \pm 0.06$ \\
    NGC 1407  &1.4        &3.8           &$21.6 \pm 7.0$           &$608.9 \pm 97.9$     &$1.26 \pm 0.07$   &$0.88 \pm 0.05$  &$1.26 \pm 0.09$ \\
    NGC 4472  &2.7        &4.9           &$14.1 \pm 1.0$           &$491.8 \pm 17.0$     &$1.19 \pm 0.02$   &$1.02 \pm 0.02$  &$1.59 \pm 0.01$ \\
    NGC 4486  &16.2	  &28.9          &$15.7 \pm 0.6$           &$520.3 \pm 9.3$      &$1.36 \pm 0.01$   &$1.60 \pm 0.04$  &$1.37 \pm 0.01$ \\
    NGC 4649  &3.0        &9.4           &$16.4 \pm 4.3$           &$530.4 \pm 69.7$     &$1.13 \pm 0.01$   &$1.08 \pm 0.01$  &$1.38 \pm 0.01$ \\
    NGC 5846  &2.3        &3.7           &$12.5 \pm 1.9$           &$463.2 \pm 35.0$     &$1.17 \pm 0.05$   &$0.88 \pm 0.05$  &$2.16 \pm 0.02$ \\

    \hline
  \end{tabular}
  \caption{Derived properties of the sample of six X-ray bright elliptical galaxies: (1) Galaxy name, (2) $\chi^2$ per data point between deprojected temperature and model temperature, (3) $\chi^2$ per data point between deprojected logarithmic pressure and model logarithmic pressure, (4) enclosed mass at 25 kpc and associated 95\% confidence range, (5) circular velocity at 25 kpc and associated 95\% confidence range, (6) power-law index of mass profiles on a log-log scale and associated 1-$\sigma$ error, (7) power-law index of mass profiles on a log-log scale fitting only until 10 kpc and associated 1-$\sigma$ error and (8) power-law index of mass profiles on a log-log scale fitting only above 10 kpc and associated 1-$\sigma$ error. \label{tab:results}}
\end{table*}

\subsection{Choice of optimal smoothing parameter}\label{sub:smooth}

We use our method in the non-Markov mode to obtain the best-fit
circular velocity curve for different values of $\lambda$ from the
pseudo deprojected temperature and density profiles. We find that
$n_{\textrm{iter}} = 10^6$ iterations is sufficient to reach a best-fit model.

A plot of the root-mean squared deviation $R$ between the recovered
circular velocity curve of the test model and the input circular
velocity curve, against the smoothing parameter $\lambda$, is shown in
Figure \ref{fig:smooth} where:

\begin{equation}
R = \left[\frac{\sum (V_{\textrm{c,test}} - V_{\textrm{c,mod}})^2}{n_{\textrm{mod}}}\right]^{1/2}
\end{equation}

\noindent Figure \ref{fig:smooth} shows that $R$ has a minimum at
around $\lambda_{opt} = 0.5$. A lower smoothing leads to unphysical
fluctuations in the recovered model and a higher smoothing results in
a recovered model that does not fit the temperature and density
profiles of the test model well.

Figure \ref{fig:test_model} shows the resulting best-fit temperature,
density and circular velocity curves for the minimum (0.001), maximum
(100) and optimal (0.5) values for $\lambda$.

\subsection{Confidence range on recovered circular velocity curve}

To obtain the range of circular velocity curves within some confidence
level, we have to ensure that our method is correctly probing the
shape of the posterior probability. Assuming the optimal smoothing
parameter found above, we now use the Markov mode on the pseudo
profiles. $n_{\textrm{tune}} = 2000$ results in an acceptance rate of
about 23\% in the subsequent proposed changes. We use a higher
$n_{\textrm{iter}} = 2 \times 10^6$, above which the profiles change
insignificantly and choose $n_{\textrm{burnin}} = 5 \times 10^5$, above which
the expectation values calculated change insignificantly. In Figure
\ref{fig:marg_post} we have plotted how the circular velocity points
at each radius are distributed in the Markov chain (i.e. after the
burn-in phase), to illustrate the difference between using a
$n_{\textrm{iter}}$ that is too low and one that is about right. In
the left plot where $n_{\textrm{iter}} = 10^4$ and
$n_{\textrm{burnin}} = 5000$, the distribution of circular velocity
points at each radius (i.e. the marginal posterior probability of the
circular velocity curve at each grid point) have not settled to a
smooth distribution, and therefore one would suspect that the chain is
not yet properly sampling the posterior probability distribution. The
plot on the right however shows that running the Markov chain for
longer results in smooth marginal posterior probability distributions
that are well described by Gaussians (illustrated by the green lines).

From Figure \ref{fig:marg_post}(b) the median value for the circular
velocity at each radius is the circular velocity for which the
marginal posterior probability is highest. The expectation value is
the mean of all the circular velocity values chosen at that radius and
the associated $c\%$ confidence range is given by the extrema of $c\%$
of the circular velocity values with the highest marginal posterior
probabilities. Figure \ref{fig:bayes} shows the results of calculating
these for all the model profiles. One can see that the median and
expected profiles lie on top of each other, showing that the marginal
posterior probability distributions at each radius are symmetrical. It
can also be seen that the test model profiles lie within or on the
boundary of the 68\% confidence range. We repeat this for different
sets of pseudo deprojected density and temperature profiles and we are
convinced that the procedure is able to recover the circular velocity
curve without a systematic bias.

\begin{figure*}
  \centering
  \subfigure{
    \includegraphics[width=5.3cm]{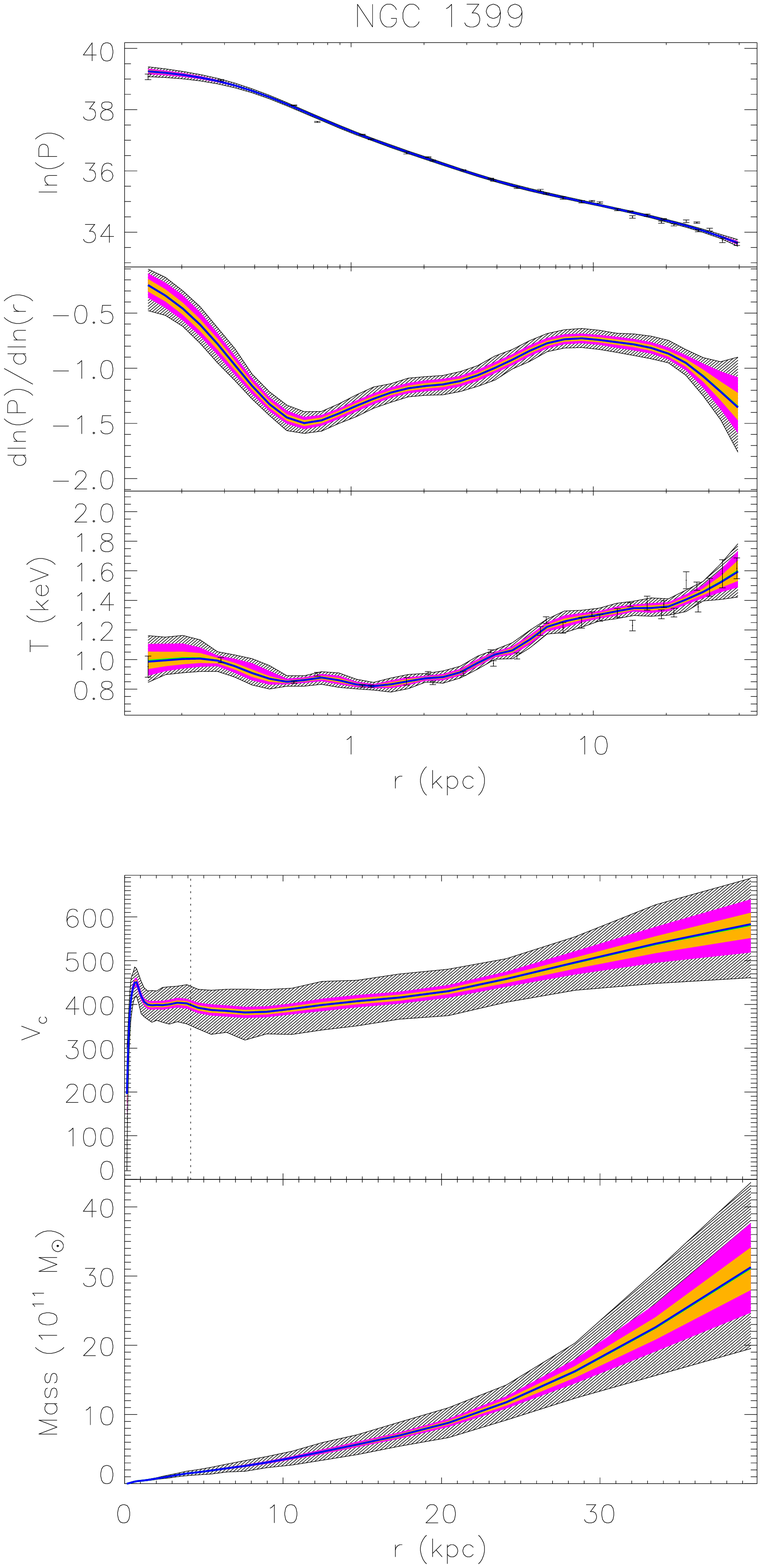}
  }
  \subfigure{
    \includegraphics[width=5.3cm]{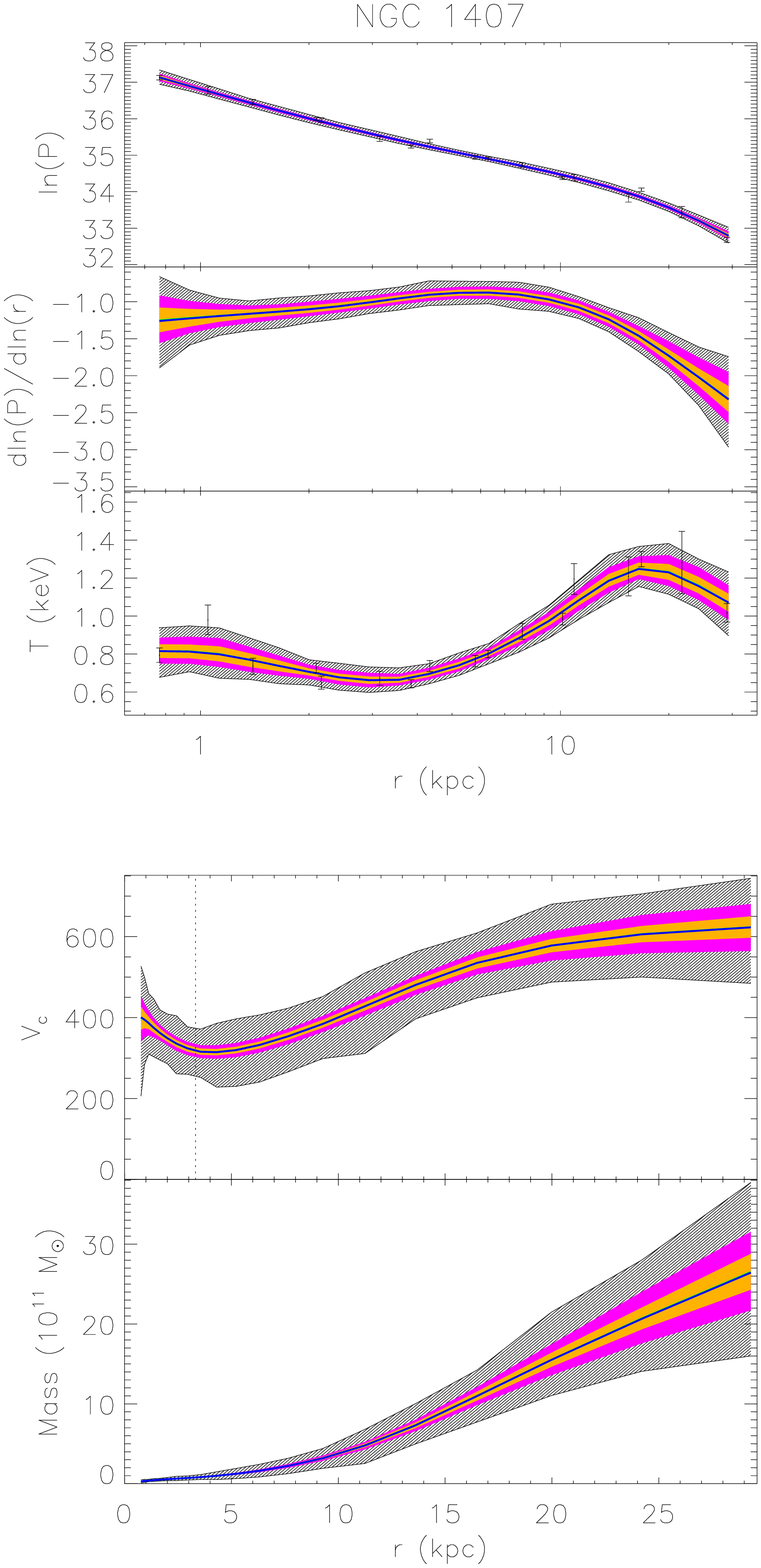}
  }
  \subfigure{
    \includegraphics[width=5.3cm]{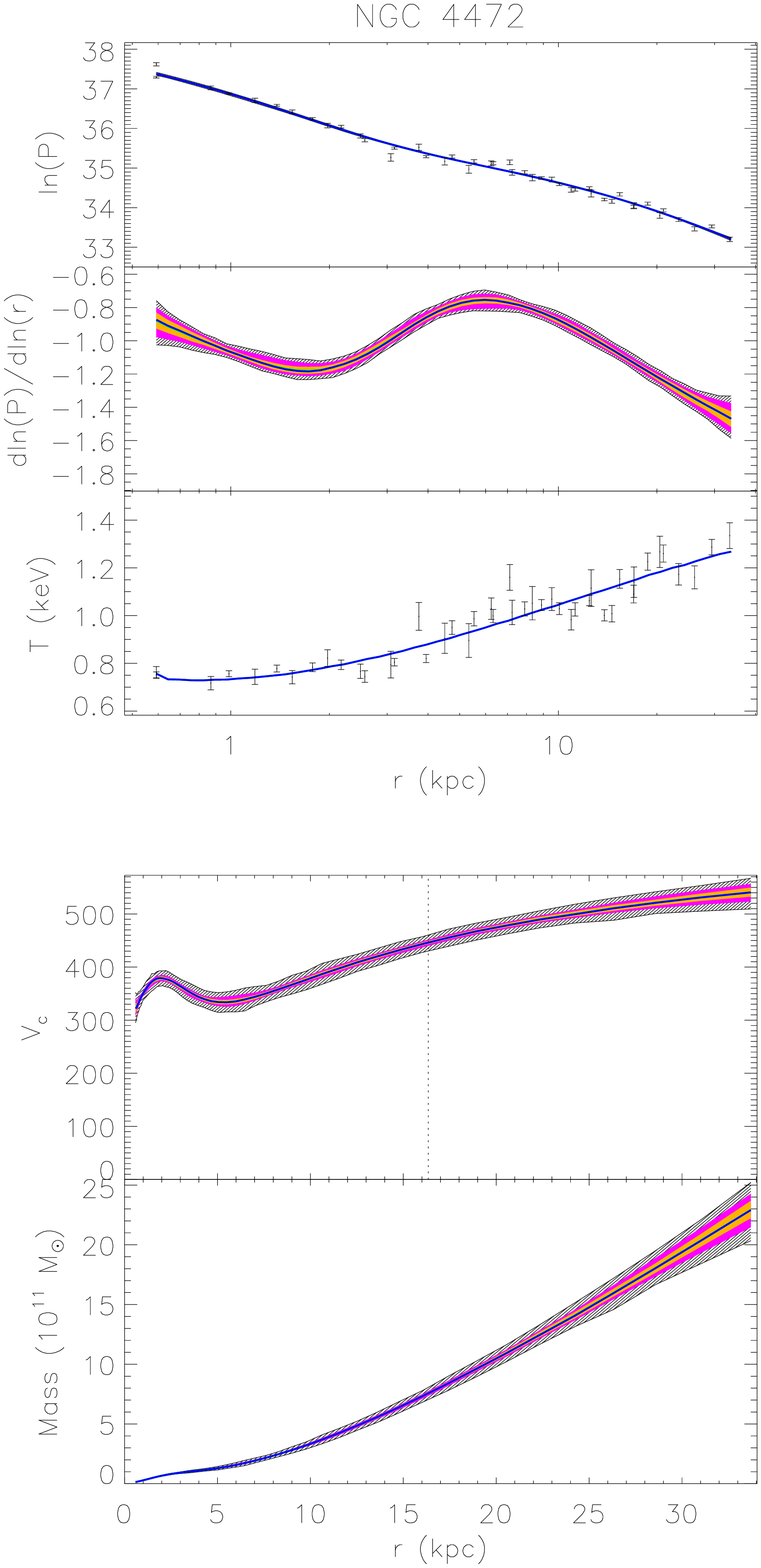}
  }
  \subfigure{
    \includegraphics[width=5.3cm]{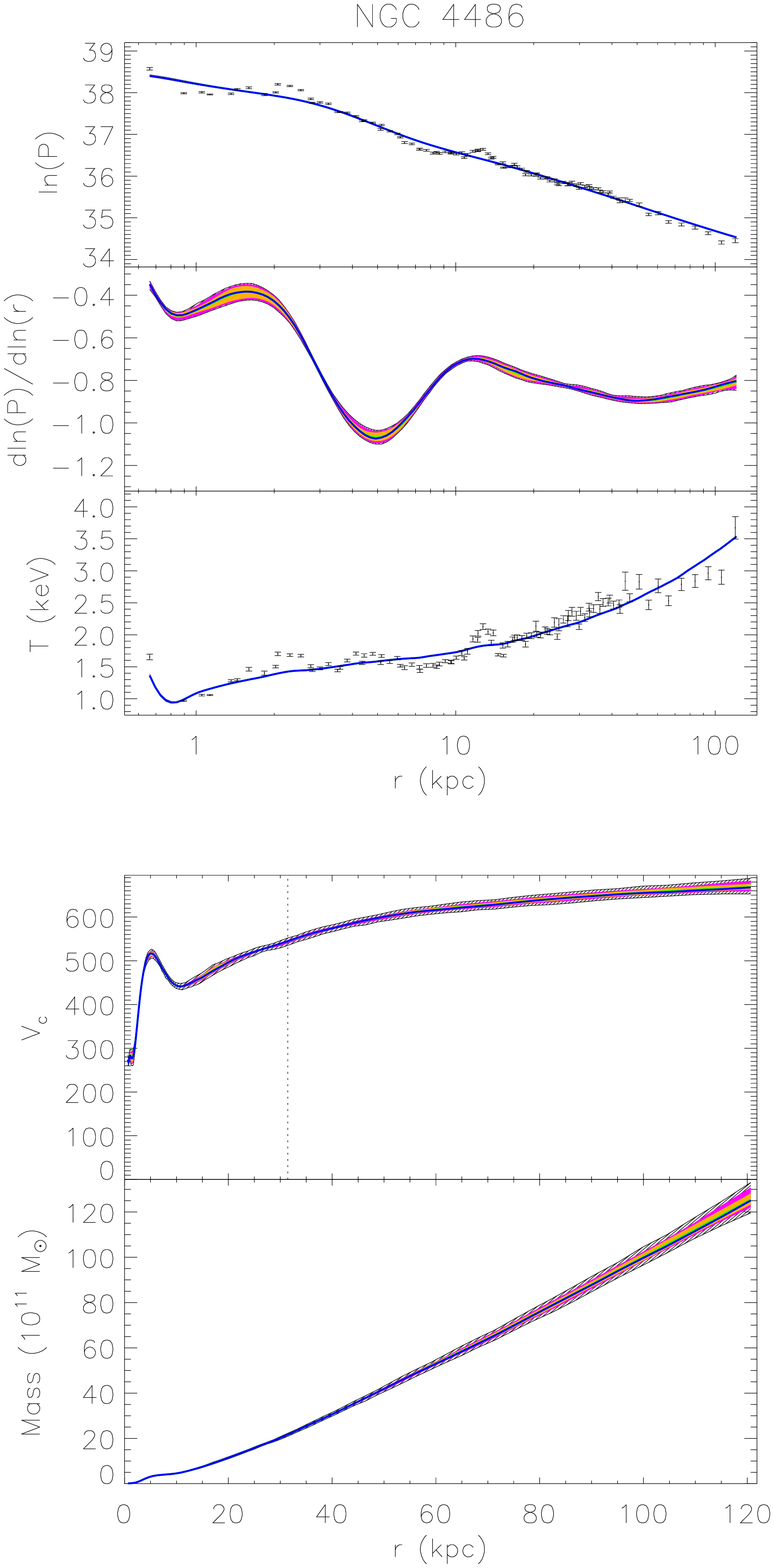}
  }
  \subfigure{
    \includegraphics[width=5.3cm]{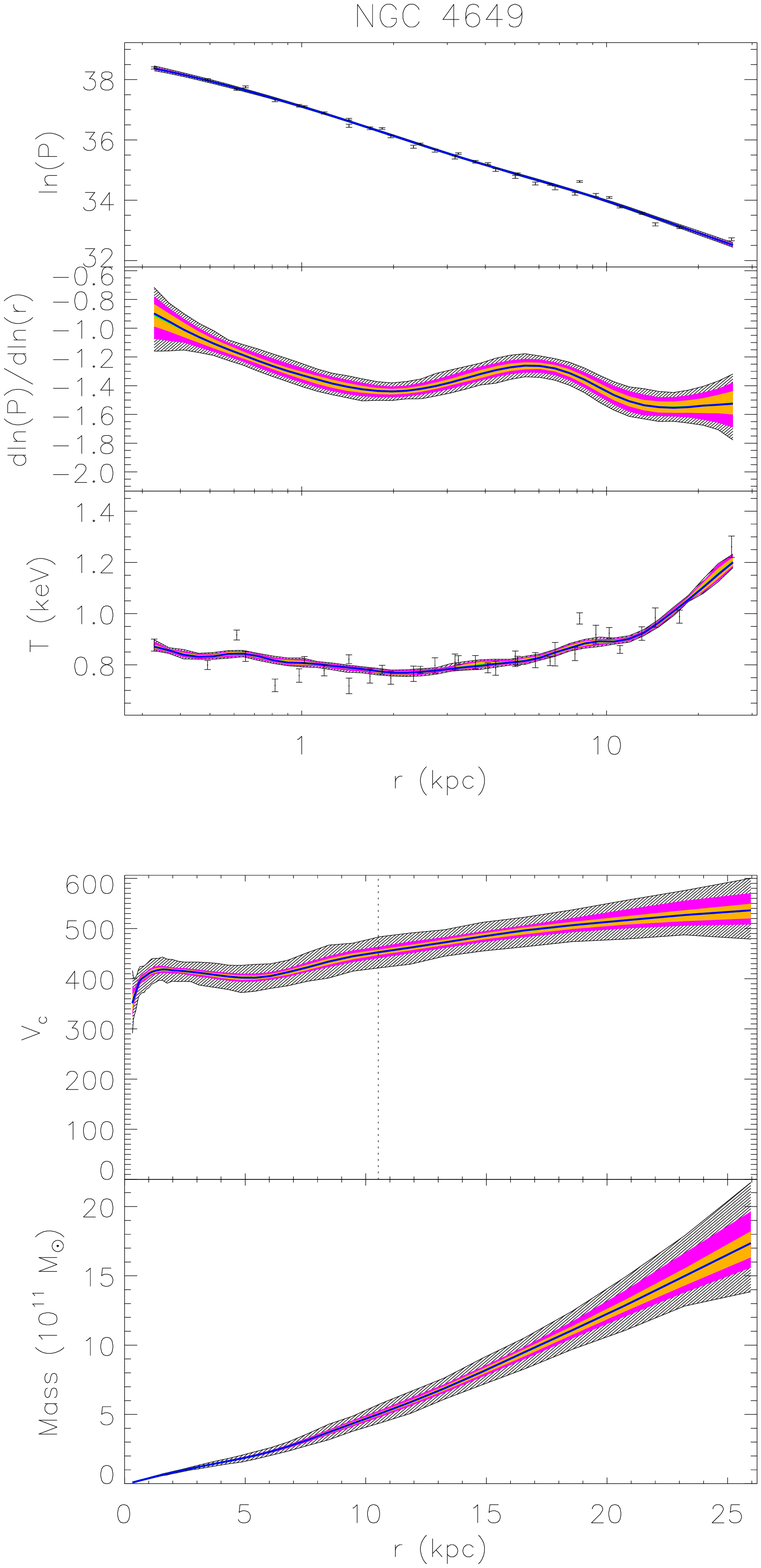}
  }
  \subfigure{
    \includegraphics[width=5.3cm]{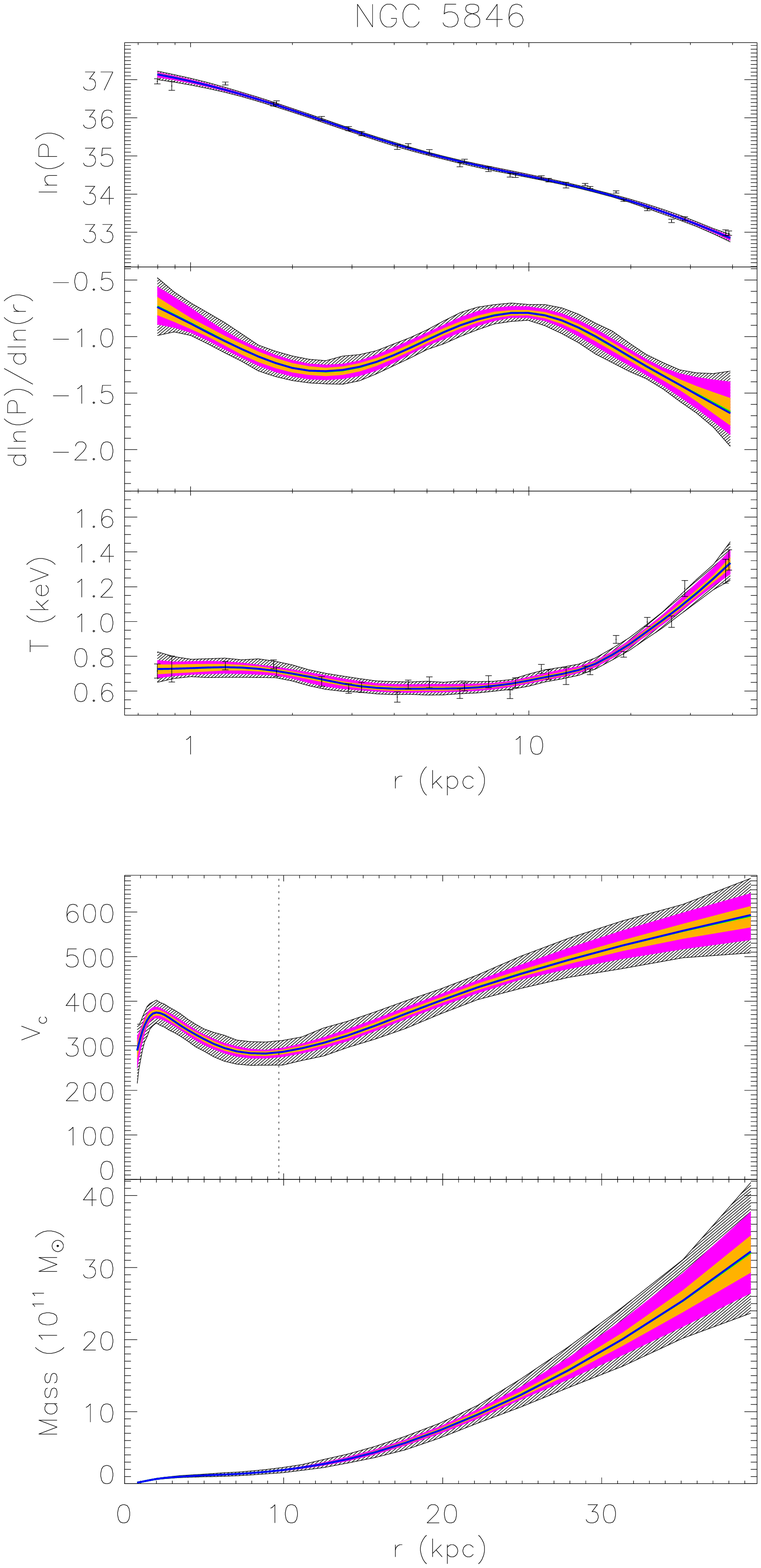}
  }
  \caption{Application of the non-parametric analysis to the sample of
    six X-ray bright elliptical galaxies: Going from top down for each
    galaxy are shown the (i) logarithmic pressure, (ii) logarithmic
    pressure gradient, (iii) temperature, (iv) circular velocity and
    (v) mass profiles.  The top three profiles are on a logarithmic
    radial scale, while the bottom two are on a linear radial
    scale. Black points on the logarithmic pressure and temperature
    plots show the deprojected profiles and associated statistical
    errors. The green and blue profiles (often indistinguishable)
    correspond to the expected and median profiles respectively. The
    grey region shows all generated points, the pink region shows the
    95\% confidence range and the orange region shows the 68\%
    confidence range. The vertical black dotted lines on the circular
    velocity plots are at $1R_e$, given in Table
    \ref{tab:stell}. \label{fig:npm}}
\end{figure*}

\section{TOTAL, STELLAR AND DARK MATTER MASS PROFILES}\label{sec:app}

Now that we have tested the method, we can apply it to the sample of
six X-ray bright elliptical galaxies to derive total mass and circular
velocity profiles (Figure \ref{fig:npm} and Table
\ref{tab:results}). We then estimate the stellar contribution to the
total mass profiles using optical photometric data and stellar
population models. Assuming negligible gas mass, we subtract the
stellar mass from the total mass profiles to infer the dark matter
mass fractions.

\begin{table*}
  \centering
  \begin{tabular}{c c c c c c c c c}
    \hline
    \hline
    Galaxy    &Source             &Band	                &Seeing ('')   &$R_e$ (kpc)    &Source     &$R_2$ ('', kpc)      &$M/L \, (M_{\odot}/L_{\odot})$  &Source\\
    (1)       &(2)                &(3)                  &(4)           &(5)            &(6)        &(7)                      &(8)                         &(9)\\                                     
    \hline                                                                                                                                                           
    NGC 1399  &S00                &$B$     		&0.05          &4.17           &K10        &2.9, 0.28                &9.7                         &Kr00\\                          
    NGC 1407  &Sp08               &$B$     		&1             &3.32           &H09        &2.7, 0.38                &4.2                         &Z07\\ 		          
    NGC 4472  &K09                &$V$     		&0.05          &16.34          &H09        &3.5, 0.28                &6.0                         &T00\\                          
    NGC 4486  &K09                &$V$		        &0.05          &31.44          &H09        &7.6, 0.59                &7.0                         &G09\\                           
    NGC 4649  &K09                &$V$     		&0.05          &10.52          &H09        &4.2, 0.34                &7.8                         &T00\\                          
    NGC 5846  &Kr00               &$V$		        &0.2           &9.70           &Kr00       &3.3, 0.39                &7.2                         &T00\\                          
    \hline
  \end{tabular}
  \caption{Photometric data for the sample of six X-ray bright elliptical galaxies: (1) Galaxy name, (2) source of surface-brightness profiles, (3) band of photometry, (4) seeing values , (5) effective radii, (6) source of effective radii, , (7) radius at which the optical surface-brightness profile has a gradient of 2, (8) stellar population mass-to-light ratios converted to the photometric band in column (3) and distances from Table \ref{tab:sample}, and (9) source of stellar population mass-to-light ratios. References are S00 \protect\citep{sag+00}, Sp08 \protect\citep{spol+08}, K09 \protect\citep{korm+09}, Kr00 \protect\citep{kron+00}, K10 (J. Kormendy, {\it private communication}),  H09 \protect\citep{hop+09}, Z07 \protect\citep{zhang+07}, T00 \protect\citep{trag+00}, G09 \protect\citep{geb+09}. \label{tab:stell}}
\end{table*}

\subsection{Total mass profiles and circular velocity curves}\label{sub:total}

The fits to the deprojected logarithmic pressure and temperature
profiles and the derived logarithmic pressure gradient, circular
velocity and mass profiles are shown in Figure \ref{fig:npm}, for the
sample of six galaxies. Columns (2)--(7) in Table \ref{tab:results}
list the $\chi^2$ per data point of the expected models, the enclosed
mass and associated 95\% confidence range at 25 kpc and the circular
velocity and associated 95\% confidence range at 25 kpc. We choose 25
kpc because the deprojected profiles for all the galaxies extend at
least as far as this radius. We quote the $\chi^2$ per data point
rather than the $\chi^2$ per degrees of freedom because the latter is
unknown. The degrees of freedom is equal to the number of constraints
subtracted by the number of parameters. In our case the number of
constraints is the number of data points plus the number of
constraints introduced by the smoothing, which is difficult to define.

Generally, the fits in the mean are very good. The $\chi^2$ per data
point between the deprojected temperature and model temperature ranges
from 1.4 to 16.2.  The $\chi^2$ per data point between the deprojected
logarithmic pressure and model logarithmic pressure ranges from 3.7
and 28.9. The $\chi^2$ values are not comparable to the usual reduced
$\chi^2$ which is $\sim 1$ at the 1-$\sigma$ level, because the
reduced $\chi^2$ is normalised by the number of degrees of
freedom. Our $\chi^2$ values may also be higher than expected because
we have only considered statistical errors on the deprojected
temperature and pressure profiles. There is also
correlated/anti-correlated scatter in the deprojected temperature
profiles between adjacent points as a result of the deprojection
procedure \citep{chur+08}. AGN activity driving shock waves into the
ICM \citep[e.g.][]{form+07} can also result in systematic undulations
in the measured temperature and density profiles. The models do not
fit the deviations caused by the correlated/anti-correlated errors and
by systematic errors because we have calibrated the smoothing to a
test model of an ideal massive elliptical galaxy.  If the statistics
of the data are low as in the case of NGC 1407, then the error bars
cover the correlated/anti-correlated scatter and the $\chi^2$ values
are low. The outstanding statistics ($\sim 0.5$ Ms of Chandra
observation) and very high surface brightness of NGC 4486 enables
deprojected temperature and density profiles to be obtained in much
narrower bins than for the other galaxies. This allows systematic
deviations to be resolved, but as the errors bars are small, they do
not cover the correlated/anti-correlated scatter or the systematic
deviations. As a result the $\chi^2$ values for NGC 4486 are on
average about 5 times higher than for the other galaxies.

The general shape of the derived circular velocity curves can be
summarised by a steep rise in the centre (except in NGC 1407) followed
by a slight dip at a few to 10 kpc and then a more gentle rise
outwards, reaching circular velocities of 463--609 km/s at 25 kpc. The
outward rise in the circular velocity curves is a combined result of
temperature profiles that rise outwards {\it and} logarithmic pressure
gradients that increase in magnitude outwards. The dips in the
circular velocity curves are a result of dips in the logarithmic
pressure gradient around the same radius. These dips may reflect true
changes in the mass distribution or they may be a result of shocks in
the gas propagating outwards. Shock waves can be produced by an
unsteady outflow of relativistic plasma from a central black hole, as
in NGC 4486 \citep{chur+08}. The shockfront is characterised by a
sharp increase in the pressure inwards and then a decrease in the
rarefaction region behind the shock. This manifests itself as a dip in
the circular velocity curve at the position of the shockfront in NGC
4486.

The enclosed mass profiles of all galaxies increase outwards with an
increasing gradient as expected from the increasing circular velocity
curves, pointing towards non-isothermal mass profiles. The 95\%
confidence ranges in the circular velocities and enclosed masses at 25
kpc are reasonably small and are smallest for the galaxies with the
best statistics (NGC 4472 and NGC 4486) because the $\chi^2$ values
are higher as discussed above.

\begin{figure*}
\centering
\subfigure[]{
  \includegraphics[width=8.5cm]{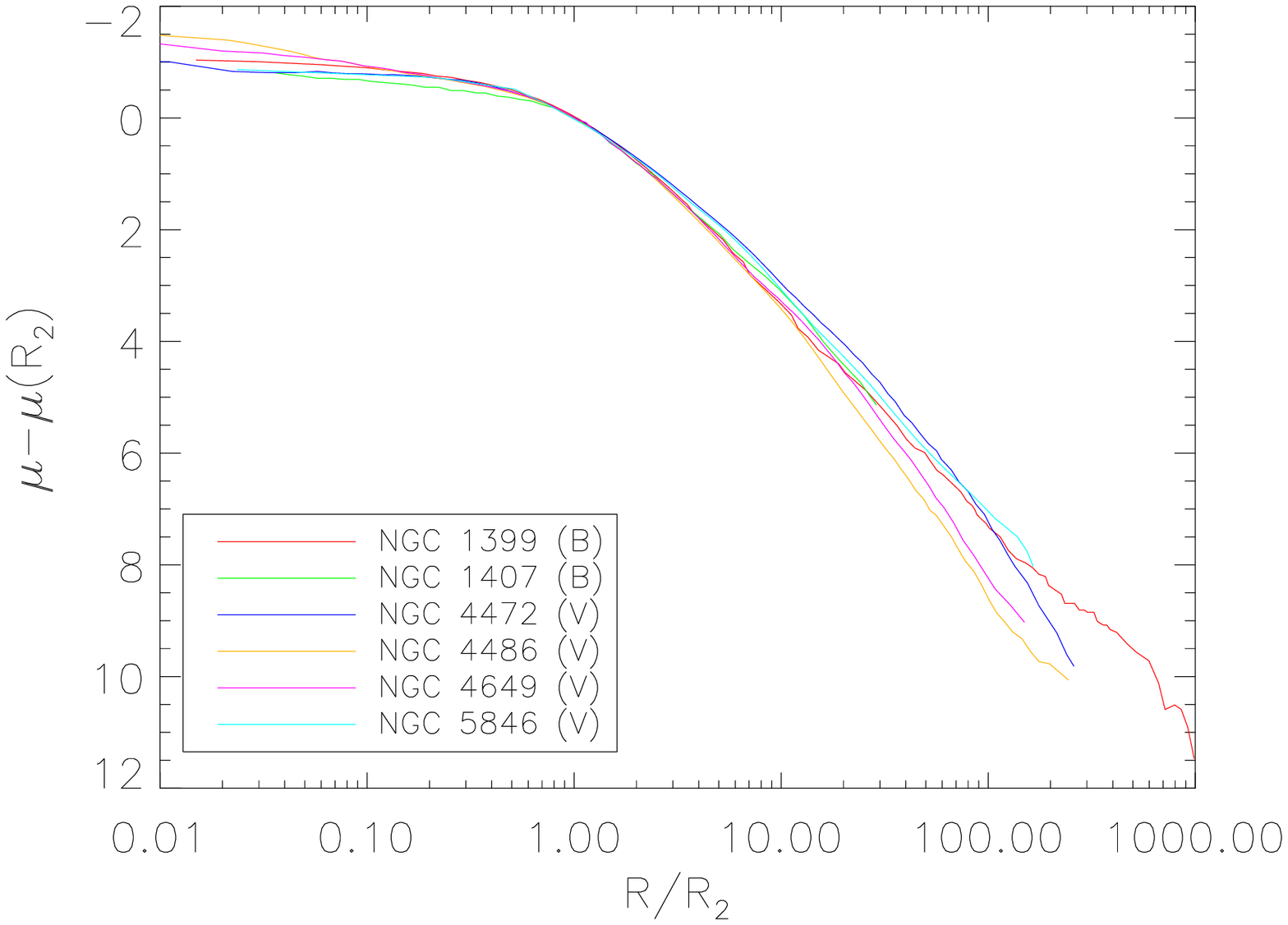}
}
\subfigure[]{
  \includegraphics[width=8.5cm]{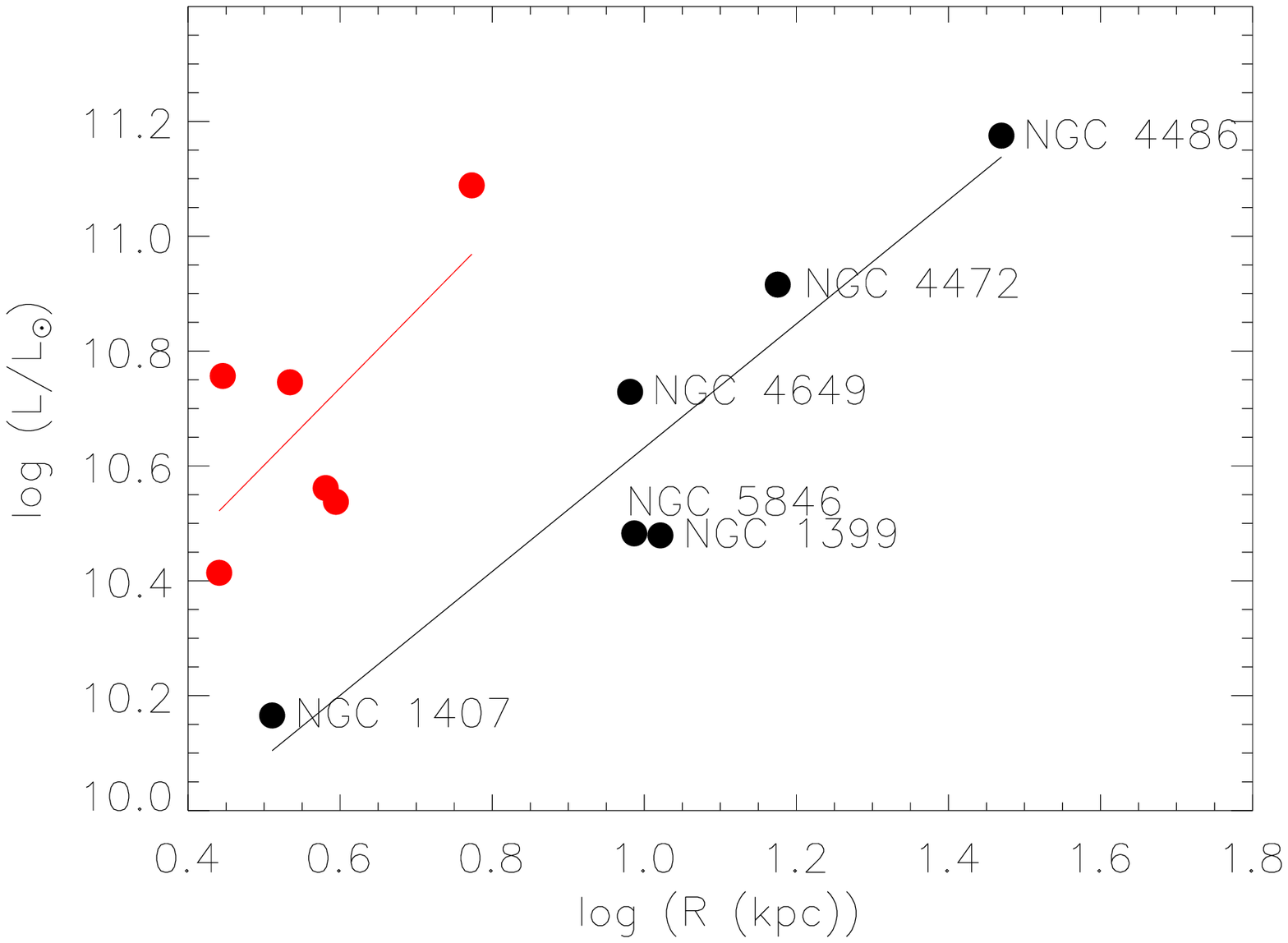}
}
\caption{Luminous properties of the six X-ray bright elliptical
  galaxies: (a) Surface-brightness profiles scaled by $R_2$ (the
  radius where $d\mu /d\log R = 2$) and by the surface brightness
    $\mu(R_2)$. (b) Kormendy relation between the $B$-band luminosity
    $L_B(30R_2$) and $R_2$ (red filled circles) and between the
    $B$-band luminosity $L_B(R_e)$ and $R_e$ (black filled
    circles). $\log(R_2)$ has been shifted by 1 to the right. The
    straight red and black lines show the respective best-fit straight
    lines.\label{fig:stell}}
\end{figure*}

\subsection{The stellar mass contribution}\label{sub:stell}

To estimate the stellar contribution to the total circular velocity
curve, we use surface-brightness profiles from various sources in the
literature, given in column (2) in Table \ref{tab:stell} and shown in
Figure \ref{fig:stell}(a). We deproject the surface-brightness
profiles assuming a spherical stellar distribution and a power law
outside the range of the data. We then integrate the resulting 3-D
luminosity density profile to obtain the luminosity profile of the
stars. Assuming constant stellar mass-to-light ratios with radius
(columns (8) and (9) in Table \ref{tab:stell}), we obtain stellar mass
profiles and stellar circular velocity curves from the luminosity
profiles.

In order to find a scaling radius that best represents the luminosity
of the stellar component, we examine two alternatives. The first is
the mean effective radius, given in Table \ref{tab:stell} for the
sample of six galaxies. For NGC 1399 we use the determination of
J. Kormendy ({\it private communication}) based on photometry in
\cite{caon+94} and \cite{lauer+07}. We calculate the mean effective
radii from the major-axis effective radii and ellipticities for NGC
1407, NGC 4472, NGC 4486 and NGC 4649 determined by \cite{hop+09} from
photometry in \cite{lauer+07} (NGC 1399), \cite{bender+88} (NGC 1407)
and \cite{korm+09} (NGC 4472, NGC 4486 and NGC 4649). \cite{hop+09}
determined double-S\`ersic fits to the profiles but calculated the
effective radii for the total profiles. We do not use their
determination for NGC 1399 because it is based on photometry from
\cite{lauer+07}, which only probes the central region. For NGC 5846,
we use the mean de Vaucouleur's effective radius from
\cite{faber+89}. The effective radii of massive elliptical galaxies
are large and therefore require extended surface-brightness profiles
with excellent sky subtraction, which we believe is true in the case
of the photometry of \cite{korm+09}, but not in the older photometry
where the sky subtraction may not be so accurate. Therefore as an
alternative, we also calculate $R_2$ (column (6) in Table
\ref{tab:stell}), the radius at which $d\mu/d\log R = 2$, where $\mu$
is in magnitudes/arcsec$^2$. Figure \ref{fig:stell}(a) illustrates the
similarity between the general shapes of the surface-brightness
profiles when the radius is scaled by $R_2$, and the surface
brightness by its value at $R_2$. We obtain values of $R_2$ ranging
between 2.7--7.6'', corresponding to 0.3--0.6 kpc, just outside
typical seeing values of 1--2''. Figure \ref{fig:stell}(a) shows that
$R_2$ is located around the radius that separates the core where the
surface brightness decays gently, from the region where it starts
decaying more steeply, and therefore may be mostly reflecting the core
properties.

To determine which scaling radius should be used as representative of
the galaxy's luminosity, we plot an analogy to the Kormendy relation
\citep{korm77} in Figure \ref{fig:stell}(b). We plot the $B$-band
luminosity of the sample galaxies at 30$R_2$ against $R_2$ and the
$B$-band luminosity at $1R_e$ against $R_e$ and fit straight lines
through both.  The best-fit straight lines have slopes of 0.74 and
0.93 and Pearson product moment correlation coefficient (PMCC) values
of 0.70 and 0.94 respectively, supporting correlations at about 88\%
and 99\% levels of significance for a sample size of six. The
correlation between the effective radius and the luminosity of the
stellar component is much stronger, and therefore we will use this to
represent the size of the stellar component.

\begin{figure*}
\centering
\subfigure[]{
  \includegraphics[width=8.5cm]{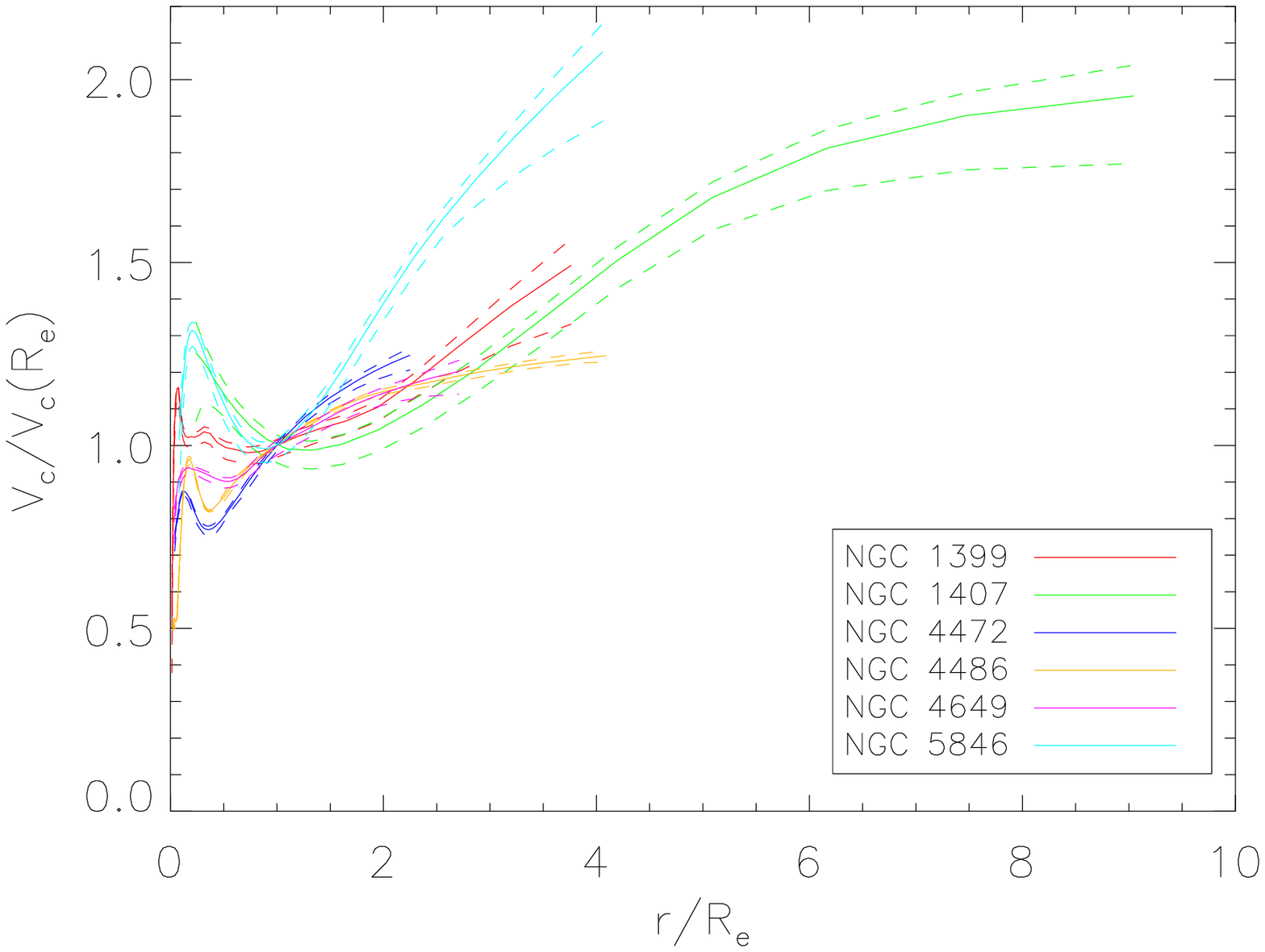}
}
\subfigure[]{
  \includegraphics[width=8.5cm]{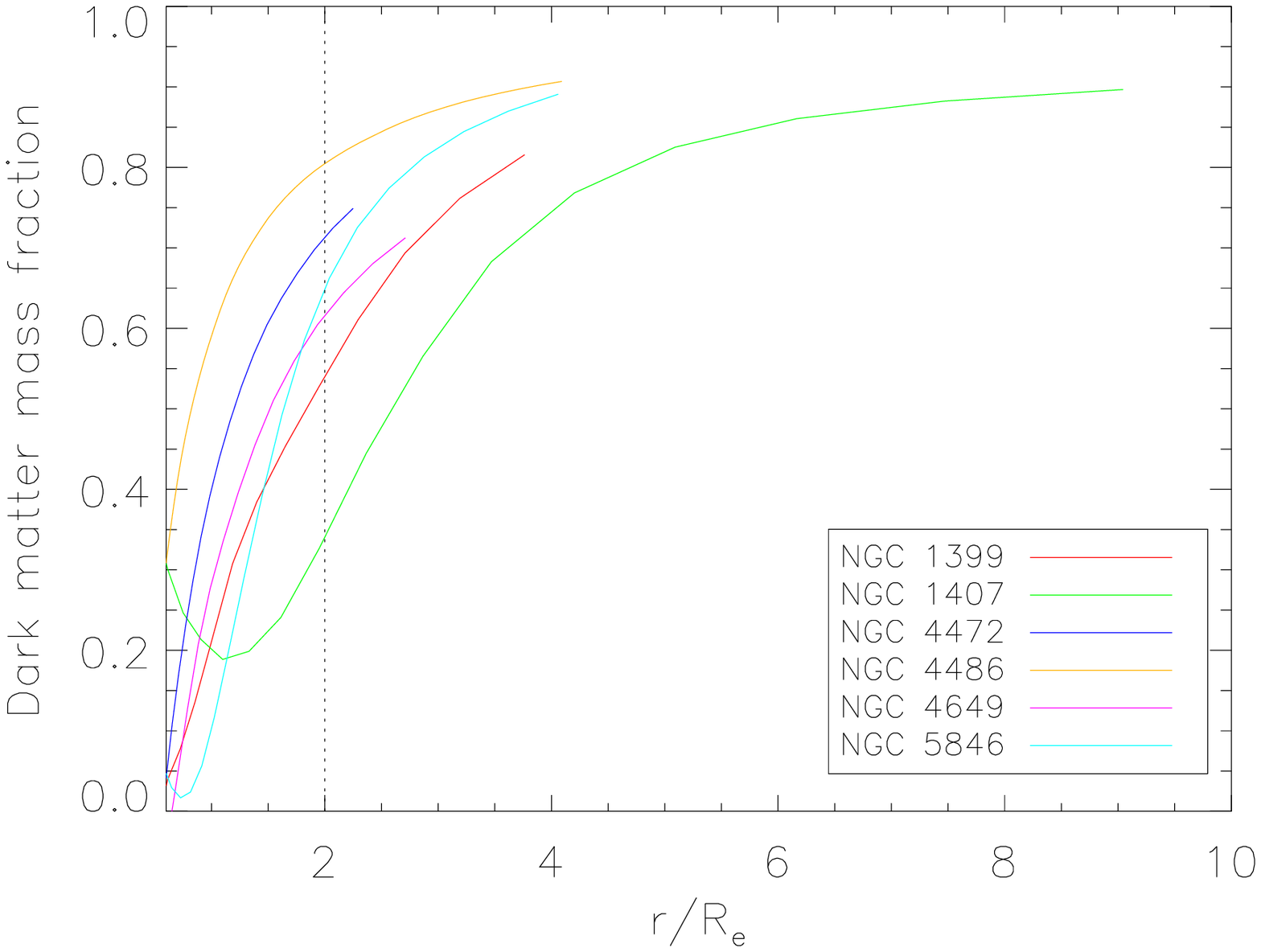}
}
\caption{Total and dark matter mass properties of the six X-ray bright
  elliptical galaxies: (a) Total circular velocity profiles scaled by
  $R_e$ in radius and the total circular velocity $V_c(R_e)$. (b) Dark
  matter mass fractions scaled by $R_e$ in radius. The black dotted
  line highlights the dark matter mass fractions at
  $2R_e$. \label{fig:dm}}
\end{figure*}

\subsection{Scaled total circular velocity curves and dark matter mass
  fractions}\label{sub:dm}

Figure \ref{fig:dm}(a) shows the scaled total circular velocity
profiles of the sample of X-ray bright galaxies from the X-rays along
with the 95\% confidence ranges with dashed lines. The radii are
scaled by $R_e$ and the total circular velocities are scaled by their
value at $R_e$. In the central $R_e$, all the profiles rise steeply
then fall again before rising more gently from 0.5--$1R_e$. The outer
rise in the circular velocity curves noted in Section \ref{sub:total},
is more easily compared between the galaxies in this scaled plot. The
curve of NGC 5846 rises most steeply. The curves of NGC 1399, NGC
1407, NGC 4472, NGC 4486 and NGC 4649 have a similar rise between
1--$2R_e$ but the curves of NGC 1399 and NGC 1407 keep rising
outwards, while the curve of NGC 4486 levels off outside $2R_e$.

The mass in gas calculated from integrating the gas density profiles
is $< 0.04$\% of the mass in stars in all galaxies except in NGC 4486
where it is $\sim 0.4\%$. Therefore we treat the gas mass as
negligible and assume that the baryonic mass component in the galaxy
is solely due to stars. Subtracting the stellar mass determined in
Section \ref{sub:stell} from the total mass obtained from the X-rays
gives the dark matter mass, shown as a fraction of the total mass in
Figure \ref{fig:dm}(b), outside $0.6R_e$. Below this radius the dark
matter fractions we obtain for some of the galaxies are below zero,
which could be a result of either an overestimate in the stellar
mass-to-ratios or an underestimate in the total circular velocity
curves derived from the X-rays, the latter of which we discuss in more
detail in Section \ref{sub:comp_dyn}. The dark matter fractions range
between 50--80\% at $2R_e \sim 6.5$--59 kpc in all galaxies except NGC
1407 where it is $\sim 35\%$.  Even further out, the profiles appear to
converge to a value between 80--90\%, pointing uniformly to a massive
dark matter halo. The larger range in the dark matter mass fractions
further in is indicative of dark matter mass profiles that have
different shapes. For example the dark matter mass profile is steepest
for NGC 4486, implying a more compact dark matter component.

\section{DISCUSSION}\label{sec:discuss}

In this section we compare the circular velocity curves we obtain from
our analysis to those obtained from previous X-ray determinations and
published dynamical models. We then address the issue of the
isothermality of the circular velocity curves and how this compares
with the literature. Finally we look at how the total circular
velocity curves may be correlated with properties of the stellar
components and the velocity dispersions of the environments to which
the galaxies belong.

\subsection{Comparison with previous X-ray
  determinations}\label{sub:comp_xray} 

\begin{figure*}
\centering
\includegraphics[width=19cm]{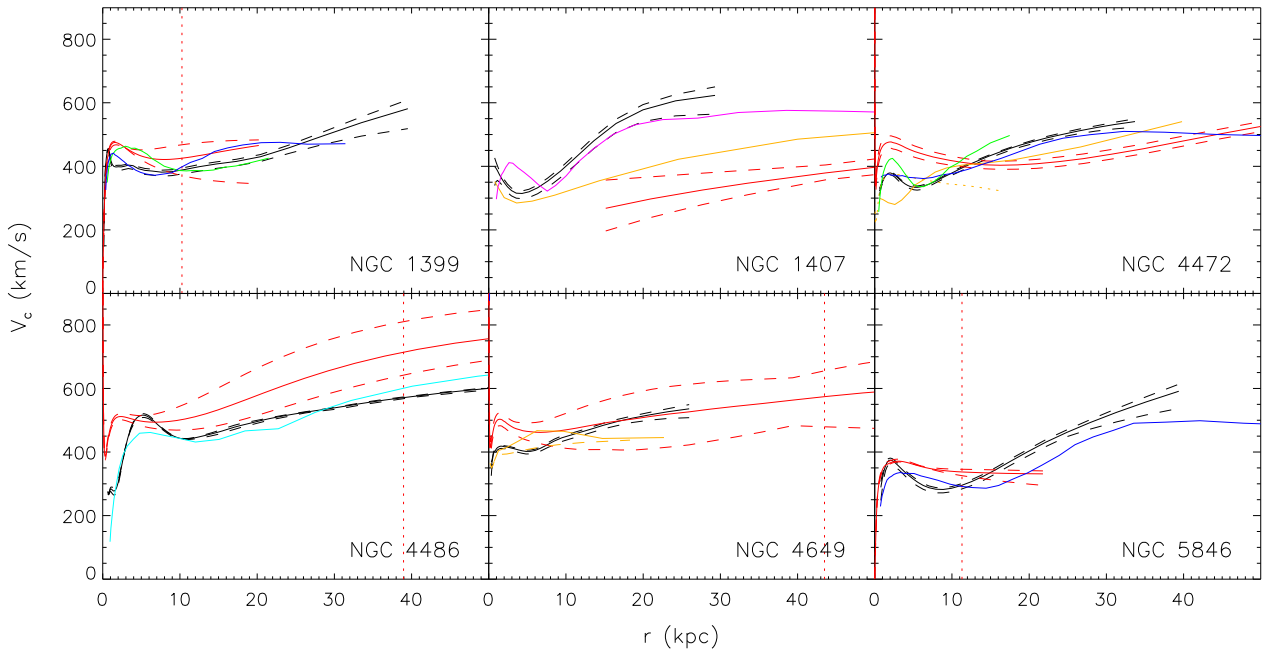}
\caption{Total circular velocity curves determined from X-rays,
  compared with those from previous X-ray determinations and published
  dynamical models: The X-ray velocity curves and associated 95\%
  confidence ranges determined in this paper are shown by the solid
  and dashed black lines respectively. Green lines \citep{fuk+06},
  blue lines \citep{nag+09}, the pink line \citep{zhang+07}, orange
  solid, dashed and dotted lines \citep{hump+06,hump+08,hump+09}, and
  the cyan line \citep{mat+02} show circular velocity curves
  determined from X-rays in the literature. The best-fit dynamical
  models along with their confidence range (for more details see text)
  are shown with solid and dashed red lines. The vertical red dotted
  line shows the radial extent of the data used in the dynamical
  modelling. \label{fig:vccomp}}
\end{figure*}

Figure \ref{fig:vccomp} shows the total circular velocity curves for
our sample of galaxies along with their 95\% confidence ranges in
black solid and dashed lines. For comparison we have overplotted
various X-ray determinations in the literature. The green lines are
from \cite{fuk+06}, who used Chandra observations of NGC 1399 and NGC
4472. The blue lines are from \cite{nag+09} who combined Chandra and
XMM-Newton observations of NGC 1399, NGC 4472 and NGC 5846, and the
pink line is from \cite{zhang+07} who combined Chandra and ROSAT data
in NGC 1407. \cite{hump+06} (solid orange lines), \cite{hump+08}
(dashed orange line) and \cite{hump+09} (dotted orange line) used
Chandra observations of NGC 1407, NGC 4472 and NGC 4649 and finally,
the cyan line is from \cite{mat+02}, who used XMM-Newton observations
of NGC 4486. Density and temperature profiles of the hot gas were
obtained from the observations and parametrised. The total circular
velocity curves were then obtained by differentiating the profiles and
applying hydrostatic equlibrium.

Discrepancies in the outer slopes are most likely a result of
differences in the spatial extent of the data used, as the last few
points anchor the outer slope. Smaller-scale differences are most
likely to arise from differences in the methods used to obtain the
mass profiles. The methods employed in the literature are all
parametric and therefore the shapes of the mass profile will have a
systematic bias. In NGC 1407 however, the discrepancy between the
outer slope of the circular velocity curve of \cite{hump+06} and that
of our determination and the determination of \cite{zhang+07} is more
signficant than between the other profiles. The source of this is
unclear.

\subsection{Comparison with dynamical models}\label{sub:comp_dyn}

Figure \ref{fig:vccomp} also shows dynamical models and their
confidence ranges from the literature, in red solid and dashed
lines. The vertical red dotted line demarcates the radial extent of
the data used in the dynamical modelling and we plot the dynamical
models to twice this radius, except in the case of NGC 1407 where we
plot the models until the radius deemed as reliable by the author. For
NGC 1399 and NGC 5846, we use the spherical dynamical models of
\cite{kron+00}. They built models by superposing spherical
distribution functions and adjusting the weights of their contribution
to the model by comparing the projection of the model to measured
surface-brightness profiles and long-slit kinematics extending to 97''
and 99'' respectively. Although this corresponds to only 5--10 kpc,
the models provide a useful comparison in the central regions. 95\%
confidence ranges are provided on the models. For NGC 1407, we compare
with the dynamical model of \cite{rom+09}, who solved spherical Jeans
equations for a varying anisotropy profile and fit to a
surface-brightness profile and globular cluster kinematics extending
out to $\sim 590$''.  They provide a 68\% confidence range on their
model. For NGC 4472, NGC 4486 and NGC 4649 we compare with the models
of Shen \& Gebhardt et al. (2010, {\it in preparation}), Gebhardt et
al. (2010, {\it in preparation}) and \cite{shen+10}, who used
axisymmetric Schwarzschild models to fit surface-brightness profiles,
integrated kinematics and globular cluster data extending to 1200'',
500'' and 533'' respectively. 68\% confidence ranges are provided on
the models.

Comparing the X-ray circular velocity curves of this paper to those
from dynamical models, we find that outside the central kpc in NGC
1399, the X-ray circular velocity curve is up to 16\% lower than the
dynamical modelling circular velocity curve. Outside a radius of about
7 kpc though, the X-ray determination lies within the 95\% confidence
range on the dynamical modelling determination. In the region of
overlap between the X-ray and dynamical circular velocity curves of
NGC 1407, the X-ray circular velocity curve rises from about 500 to
600 km/s, while the dynamical circular velocity curve rises from about
270 to 320 km/s, showing that the X-ray velocity curve is almost twice
as high throughout and therefore rises twice as steeply. The X-ray
circular velocity curve of NGC 4472 is up to 27\% lower than the
dynamical modelling circular velocity curve within about 12 kpc, but
then rises to up to 20\% higher outside this radius. Within 5 kpc, the
X-ray circular velocity curve of NGC 4486 is up to 45\% lower than the
dynamical circular velocity curve, and outside this region, the X-ray
circular velocity curve is up to 21\%. For NGC 4649, the confidence
ranges overlap from 7 kpc outwards and the circular velocity curves
agree very well outside 10 kpc. Within this radius the X-ray circular
velocity curve is at worst about 21\% lower than that of the dynamical
model. In NGC 5846, the X-ray circular velocity curve is up to 18\%
lower in the central 12 kpc than the dynamical circular velocity curve
but then rises much more steeply compared to the dynamical circular
velocity curve.

\subsubsection{Uncertainties in the X-ray analysis}\label{subsub:uncert_xrays}

Here we discuss uncertainties in the X-ray analysis, which could
explain the discrepancies we observe between the dynamical and X-ray
circular velocity curves.

One subject discussed in \cite{chur+10} is the contribution of
non-thermal components in the gas to the total pressure. The existence
of these components would lead to a systematic bias in our
determination of the mass profile derived from hydrostatic equilibrium
as we have only considered the thermal gas pressure. \cite{chur+08}
listed several non-thermal components that are consistent with
AGN-heated cores of X-ray bright elliptical galaxies exhibiting
properties of a cooling flow region: (i) Cosmic rays and magnetic
fields uniformly mixed with the thermal gas, (ii) cosmic rays and
magnetic fields forming bubbles that are free of thermal gas and (iii)
microturbulence in the thermal gas. If the fraction of the total
pressure that is due to thermal gas $f_t$ is constant with radius,
then $\phi_t = f_t \phi_{\textrm{true}}$, where $\phi_t$ is the
potential calculated from the thermal pressure and
$\phi_{\textrm{true}}$ is the true potential. This translates to $f_t
= V_{c,t}^2/V_{\textrm{c,true}}^2$. As $0 \le f_t \le 1$, non-thermal
contributions to the pressure are a possible mechanism for lowering
the circular velocity curve calculated from the X-rays and therefore
could explain the discrepancies we see in NGC 1399, NGC 4472, NGC 4649
and NGC 5846 in the central $\sim 10$ kpc and throughout NGC
4486. Calculating $f_t$ (for $r > 1$ kpc) where the X-ray circular
velocity curve is lowest compared to the dynamical circular velocity
curve gives upper limits on the local contribution of non-thermal
components to the total pressure of 30\%, 45\%, 69\%, 32\% and 33\% in
NGC 1399, NGC 4472, NGC 4649 and NGC 5846. These local, inner values
are higher than those quoted in \cite{chur+10}, where the $f_t$ is a
global estimate from fits to the total potential.

Another possible source of discrepancy could be in the assumptions
made about the abundance profile in the determination of deprojected
temperature and density profiles. \cite{chur+10} assumed a flat
abundance profile for their analysis but calculated the average effect
on the circular velocity curves if this assumption was relaxed. They
found that the circular velocity values generally increase for their
sample by a value of $\sim 2\%$. This small effect is in agreement
with \cite{werner+09}, who combined very high-resolution XMM Newton
data with Chandra data in NGC 4636 and concluded that allowing for a
varying abundance profile does not change the derived deprojected
temperature and density profiles signficantly. \cite{chur+10} also
considered the effect of the unresolved population of low-mass X-ray
binaries and estimated that this generally lowers the circular velocity
curve by $\sim 2\%$, which to some extent counteracts the effect of
allowing an abundance gradient.

We also consider the effect of the extrapolation in the deprojection
of the spectra on the derived circular velocity profiles. The
deprojected density profiles presented in \cite{chur+10} and in Figure
\ref{fig:xray} in this paper assume a power-law extrapolation based on
a power-law fit to the data. We consider two alternative
extrapolations of the Chandra density profile of NGC 4472, to
investigate whether we can explain the discrepancy between the X-ray
and dynamical circular velocity curves outside $\sim 11$ kpc. If the
index of the power law used in the extrapolation of the density
profiles becomes less negative by 0.1 (i.e. the density profile
becomes more shallow), the outermost Chandra circular velocity point
decreases by about 10 km/s, and effects further in are not
discernable. Therefore this is not a realistic explanation for the
discrepancy we observe. If instead the power-law index becomes more
negative by 0.1 (i.e. density profile drops off more steeply), then
the effect is to increase the circular velocity of the last two
Chandra circular velocity points by $\lesssim 10$ km/s. The effects
further in are negligible. This would not be sufficient to explain why
the X-ray circular velocity curve is so much higher in NGC 1407
compared to the dynamical circular velocity curve.

Shocks in the gas result in a local violation of hydrostatic
equilibrium and in spatial correlations in the pressure and circular
velocity curves (Section \ref{sub:total}). In NGC 4486, \cite{chur+08}
showed a simulated shock at 2.7 arcmin ($\sim 13$ kpc), which could
correspond to the acute bump seen in the pressure and temperature
profiles around this radius in Figure \ref{fig:npm}. They found that
such a shock leads to a dip in the potential of NGC 4486. We observe a
similar dip in the total circular velocity of NGC 4486 close to this
radius. In the other galaxies, shockfronts are not obvious in the
deprojected profiles, possibly due to the lower statistics compared to
NGC 4486. In general, small-scale deviations from hydrostatic
equilibrium are simply not fit due to the smoothing prior in our
method, while a larger-scale violation results in a systematic effect.

\cite{chur+10} also discussed the effects of assuming spherical
symmetry in their determination of the gravitational potential from
X-rays and found that on average the effect is unlikely to exceed
7\%. As the effect is systematic and will either increase/decrease the
average slope of the potential, the resulting change in the circular
velocity is of the order of 4\%.

In earlier work, \cite{chur+08} estimated biases introduced when
fitting multi-temperature plasma with a single-temperature model by
comparing profiles obtained in different energy bands.  In the case of
NGC 4486, they compared temperature and density profiles calculated
from broad-band spectra to that from hard-band spectra and found that
differences are small outside 0.4 arcmin, which is a couple of kpc.
Within this region however, AGN activity can result in hot and cool
gas structures superposed along the line of sight. NGC 4486 is the
most disturbed galaxy in our sample but this source of uncertainty
could also explain some of the discrepancies we find in the central
few kpc in other galaxies in the sample.

\subsubsection{Uncertainties in the dynamical modelling}\label{subsub:uncert_dyn}

Here we examine uncertainties associated with the dynamical modelling
methods, which could be the source for some of the discrepancies we
observe between the dynamical and X-ray circular velocity curves.

In the dynamical models of \cite{kron+00}, the data used extends only
to 5--10 kpc. However in the X-ray bright elliptical galaxies in our
sample, the density profiles of the stars are shallow in the outer
parts, which implies that there is a lot of stellar mass at large
radii where the orbital structure remains poorly constrained. If
additionally the orbital structure of these outer stars were very
radial, as in the numerical simulations of \cite{abadi+06} and the
dynamical models of \cite{lorenzi+08,lorenzi+09} ($\beta = 1 -
\sigma_t^2/\sigma_r^2 \sim 0.5$), there would be many stars on highly
elliptical orbits traversing large distances in the galaxy. In the
central regions they turn around and contribute to the LOS velocity
dispersion more than compared to if they were on a circular orbit
there. Therefore a higher mass profile could be inferred in the
central region, compared to what would be inferred if the outer
kinematics are also available. To estimate how much extra mass may be
inferred in the central region we look at the axisymmetric
Schwarzschild models of Coma elliptical galaxies in
\cite{thomas+07}. They used kinematic data extending to between 13--30
kpc, further out than for the dynamical models in \cite{kron+00}, and
provide 68\% confidence ranges on their circular velocity
profiles. Without the outer kinematics, they would have inferred a
less massive dark matter halo, and therefore a higher central stellar
mass-to-light ratio. Let us assume that this is approximately given by
the upper boundary of the confidence range on the mass profiles in the
central region. Therefore the inner mass profile would be
overestimated by about 50 km/s without the outer kinematic data, which
is about 10\% of the circular velocities we obtain from the
X-rays. Therefore it could partly but not wholly explain the
discrepanicies in regions where the dynamical circular velocity curves
are higher than the X-ray circular velocity curves.
  
A similar argument is relevant in cases where the dynamical models do
not have central kinematic constraints on the orbital structure, such
as in NGC 1407, where only globular cluster kinematics were used from
15 kpc outwards. Central kinematic constraints are invaluable as they
are usually of much higher quality than kinematic data further out. In
radially anisotropic systems in particular, they not only place
stringent constraints on the central dynamical structure, but also
constrain the orbital structure at least out to a few times their
maximum projected radius.

An assumption on the symmetry of the stellar distribution in the
dynamical models can also lead to a systematic bias. Let us consider
the case where the stellar distribution is assumed to be spherical but
in reality is prolate axisymmetric along the line-of-sight (LOS). Let
us position the LOS along the $z$-axis, which will then be aligned
with the long axis of the prolate ellipsoid. Therefore the velocity
dispersion along the LOS is higher than in the plane of the sky. As we
are measuring only the LOS velocity dispersions, a spherical dynamical
model would overestimate the average 3-D velocity dispersion
($\sigma_{\textrm{3D}}$) and therefore also the mass and circular
velocity. For an isotropic velocity dispersion tensor,
$\sigma_{\textrm{3D,sp}} \approx \sqrt{3} \sigma_{zz}$. If the stellar
mass distribution is in fact prolate along the LOS, then
$\sigma_{\textrm{3D,pr}} = (2\sigma^2_{xx} +
\sigma^2_{zz})^{\frac{1}{2}}$, where $x$ and $y$ are along the short
axes and $\sigma_{xx} = \sigma_{yy}$ because of axisymmetry. The {\it
  Tensor Virial Theorem} relates the intrinsic axial ratios of the
stellar mass distribution to its rotation and the anisotropy of the
velocity dispersion tensor. Assuming no rotation in this galaxy gives
$\sigma_{\textrm{3D,pr}} \approx \sigma_{zz} (1 +
2q^{0.9})^{\frac{1}{2}}$ \citep[e.g.][]{gerhard94}, where $q$ is the
intrinsic flattening. Therefore $V_{\textrm{c,true}}^2/V_c^2 \approx
\sigma_{\textrm{3D,pr}}^2/\sigma_{\textrm{3D,sh}}^2 \approx
(1+2q^{0.9})/3$.  This estimate does not take into account that
dynamical models created assuming different shapes for the galaxies
may prefer different orbit distributions, which influence the $\chi^2$
minimisation usually employed. Nevertheless, it is a useful way for us
to determine approximately how prolate a galaxy has to be along the
LOS to explain the discrepancies we observe. Looking at the radii
where the dynamical and X-ray circular velocity curves of NGC 1399,
NGC 4472, NGC 4486, NGC 4649 and NGC 5846 are most discrepant, we find
that for $r > 1$ kpc, maximum flattenings of $q = 0.52, 0.29, 0.14,
0.48, 0.47$ respectively would be required. \cite{tremb+96} found that
the distribution of intrinsic, short-to-long axis ratios peaks at
around 0.75 for bright elliptical galaxies. Therefore it is unlikely
that prolate axisymmetry along the LOS is the cause for the observed
discrepancies.

Finally, the mass profiles explored in the dynamical models are
paramerised so that that stellar mass contribution follows the shape
of the stellar luminosity profile (i.e. the mass-to-light ratio is
constant with radius). The dark matter mass contribution is usually
given by non-singular isothermal spheres or NFW profiles. This results
in a bias in the shape of the mass profiles and circular velocity
curves, especially noticeable in the central regions where the shape
of the circular velocity profile changes most.

Based on our discussion on the uncertainties inherent in the
derivation of circular velocity curves from X-rays and from dynamical
models, we conclude that the regions where the X-ray circular velocity
curves are lower compared to dynamical circular velocity curves in NGC
1399, NGC 4472 and NGC 5846 can probably be attributed to a
combination of: i) A contribution of non-thermal components to the
pressure in the X-rays, ii) multiple-temperature components in the
gas, iii) an incomplete spatial converage in the kinematic data used
in the dynamical models, and iv) insufficiently general mass profiles
in the dynamical modelling. As the dynamical models for NGC 4486 and
NGC 4649 incorporate extended kinematic data, the probable explanation
is: i) the existence of non-thermal pressure components, ii)
multiple-temperature components, and iii) insufficiently general mass
profiles in the dynamical models only. The discrepancy between the
X-ray and dynamical circular velocity curves in NGC 1407 remains
unclear.

\begin{figure}
\centering
\includegraphics[width=8.0cm]{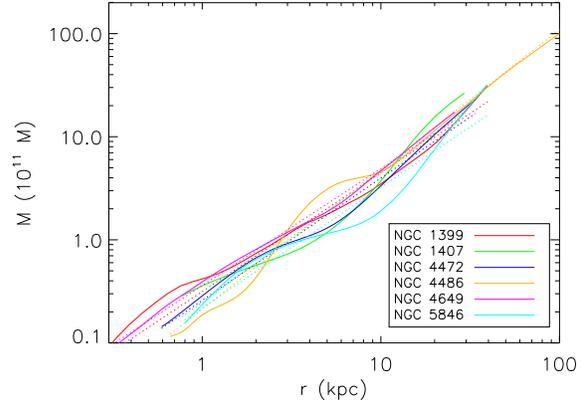}
\caption{The non-isothermality of the mass profiles of the sample of
  six X-ray bright elliptical galaxies: Solid lines show the mass
  distributions in a double-logarithmic scale. Dotted lines show
  straight lines fitted to the mass distributions in this
  scale.\label{fig:zeta}}
\end{figure}

\subsection{How non-isothermal are the mass profiles?}\label{sub:iso}

In the previous section we carried out an individual comparison
between the circular velocity curves we obtained to those obtained in
the literature. Here we examine the general property of rising
circular velocity curves noted in Section \ref{sub:total}. To quantify
how non-isothermal the circular velocity curves are, we fit a
power-law model $M \propto r^{\zeta}$ to the mass profiles as done by
\cite{fuk+06} and \cite{hump+09} from their analysis of XMM-Newton and
Chandra observations, and \cite{chur+10} on the sample we work
with. Our determinations of $\zeta$ are listed in columns (6)--(8),
calculated by fitting to the total mass profile, the mass profile
within 10 kpc and the mass profile outside 10 kpc. Figure
\ref{fig:zeta} shows the mass profiles in a log-log scale with the
best-fit models fit over the whole radial range. In this plot, the
slope of the best-fit lines is equal to the power-law index
$\zeta$. For a galaxy with an isothermal mass profile, we would expect
$\zeta=1$. Fitting to the the total mass profile we find $1.13<\zeta
<1.36$ and a mean $\langle \zeta \rangle = 1.21$. As the 1-$\sigma$
errors range between 0.01 and 0.07, the sample of these six elliptical
galaxies appear to have signficantly non-isothermal mass
profiles. Fitting the mass profiles within 10 kpc, we find $0.88<\zeta
<1.60$ and a mean $\langle \zeta \rangle = 1.10$. For NGC 1399,
$\zeta$ does not change, for NGC 4486, $\zeta$ becomes much larger,
possibly due to the shockfront at this radius, but for the remaining
galaxies $\zeta$ decreases to values between 0.88 and
1.08. Calculating $\zeta$ from the mass profiles outside 10 kpc, we
find higher values ranging from $1.26<\zeta <2.16$ and a mean $\langle
\zeta \rangle = 1.56$. This implies that the central 10 kpc of these
galaxies may be isothermal, but outside this radius the circular
velocity curves are rising. The rising circular velocity curves could
be showing that we have probed the mass distribution sufficiently far
out to observe the effects of the massive group/cluster-sized haloes
in which these galaxies reside.

\cite{fuk+06} obtained mass profiles by applying hydrostatic
equilibrium to parametrised Chandra and XMM-Newton temperature and
density profiles for a sample of 53 elliptical galaxies. They fit
power-law mass models outside 10 kpc, finding on average $\zeta = 1.33
\pm 0.33$, which overlaps with our determination of 1.56.

\cite{chur+10} found a mean $\langle\zeta\rangle = 1.11$ for NGC 1399,
NGC 4472, NGC 4486, NGC 4649 and NGC 5846, fitting their potentials
between 0.1' and 5'. Over the same radial range for the same galaxies,
we obtain $\langle\zeta\rangle = 1.17$, which is only $\sim 5$\%
higher.

\cite{hump+10} derived mass profiles from a parametric Bayesian
analysis of Chandra observations for a sample of 10 galaxies, groups
and clusters. They obtained $0.95 < \zeta <1.8$ fitting to the mass
profiles between 0.2--$10R_e$ but for the four galaxies in their
sample, they obtained $0.95<\zeta < 1.09$, i.e., close to
isothermal. We believe that the differences between their results and
ours is due to: i) Calculation of $\zeta$ over different radial ranges
from ours and ii) environmental effects, i.e. our galaxies are
located specifically at the centre of sub-groups, groups or clusters
while their galaxies, though belonging to similar environments, are
generally not at the centre. The radial range is important in the case
of NGC 4649, the only galaxy common to both samples. They used data
until about $\sim 17$ kpc, converted to our assumed distance. If we
look at Figure \ref{fig:xray} we can see that there is an extra
deprojected temperature and density point that we have from the
XMM-Newton analysis, which increases the gradient of the mass
profile. There is also reason to believe this increase because of how
similar our derived circular velocity curve is to that from
\cite{shen+10} discussed in Section \ref{sub:comp_dyn}. For NGC 1332
\cite{hump+10} used data only until $\sim 20$ kpc, which also may not
be sufficiently extended (our profiles all extend to at least $\sim
25$ kpc and in the case of NGC 4486, until almost 110 kpc.) NGC 720,
NGC 4261 (for which they have data until $\sim 31$ and $\sim 34$ kpc)
and NGC 1332 are not located at the centre of their respective
environments. Therefore even if the profiles were as radially extended
as ours, we would not expect the circular velocity curves to rise as
much as for our sample galaxies, which are all located at the centre
of their environments. For the groups and clusters in the sample of
\cite{hump+10}, they found mass profiles that increase more steeply
than isothermal mass profiles, therefore supporting the hypothesis
that we are partially probing the region where the group/cluster
potentials start dominating. The comparison between calculating
$\zeta$ within 10 kpc with calculating it over the whole radial range
of the deprojected temperature and density profiles also supports this
hypothesis.

\cite{gerhard+01} analysed the dynamical models of \cite{kron+00} and
found that their circular velocity curves are flat to within 10\% for
$R \gtrsim 0.2R_e$. These models use photometric and kinematic data
extending to less than $1R_e$ for 9/21 of the galaxies, between
1--$2R_e$ for 6/21 of the galaxies and 2--$3R_e$ for 6/21 of the
galaxies.  \cite{koop+09} solved constant-anisotropy spherical Jeans
equations for 58 strong-lens early type galaxies using stellar
velocity dispersion constraints and lensing-based total masses within
the central effective radius.  They found an average logarithmic
mass-density slope of -2.085 (for an isothermal mass profile, one
would expect -2.0) and that the dependence of this result on the value
of the anisotropy parameter $\beta$ is small.  Both studies used
elliptical galaxies that cover a range of luminosities and are not
constrained to reside at the centre of sub-group/group/cluster
environments. This means that for many of their galaxies, we have no
reason to expect a rising circular velocity curve and for those that
we do, the mass profiles may not be probing far out enough and/or the
lack of extended photometric and kinematic constraints could be
biassing the inner mass profile as discussed in Section
\ref{subsub:uncert_dyn}.

\begin{figure}
\centering
\subfigure[]
{
  \includegraphics[width=7.9cm]{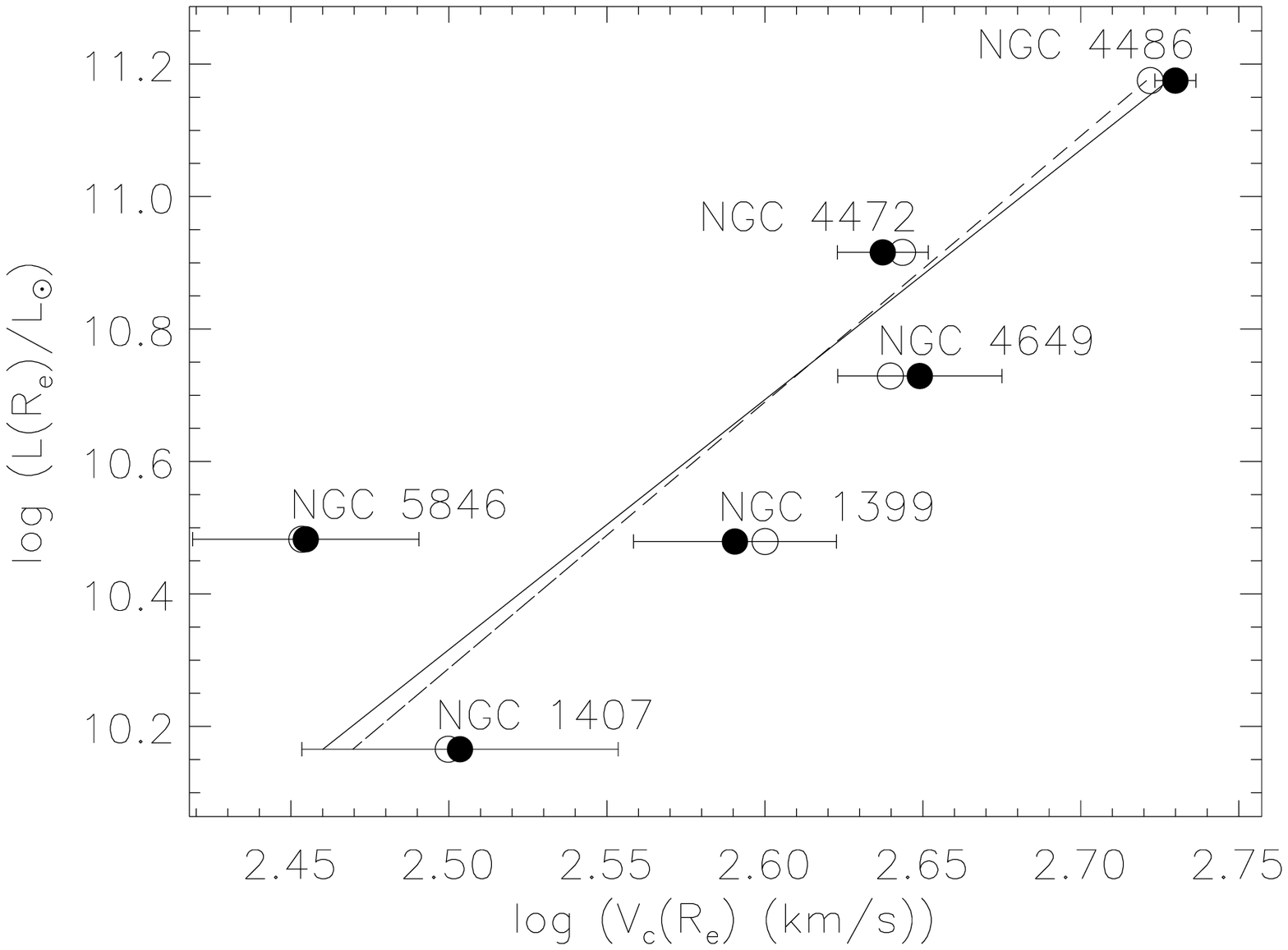}
}
\subfigure[]
{
  \includegraphics[width=7.9cm]{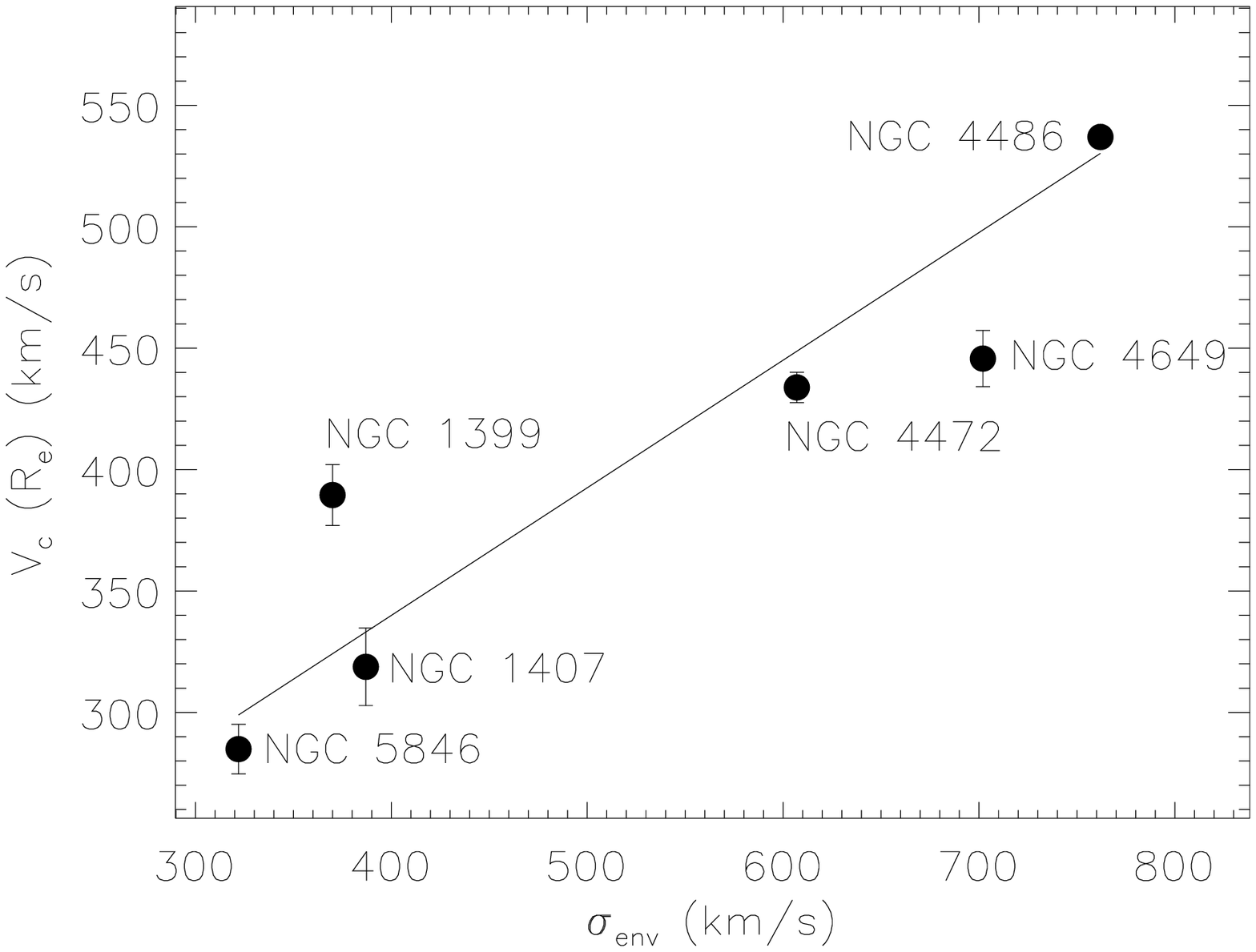}

}
\subfigure[]
{
  \includegraphics[width=7.9cm]{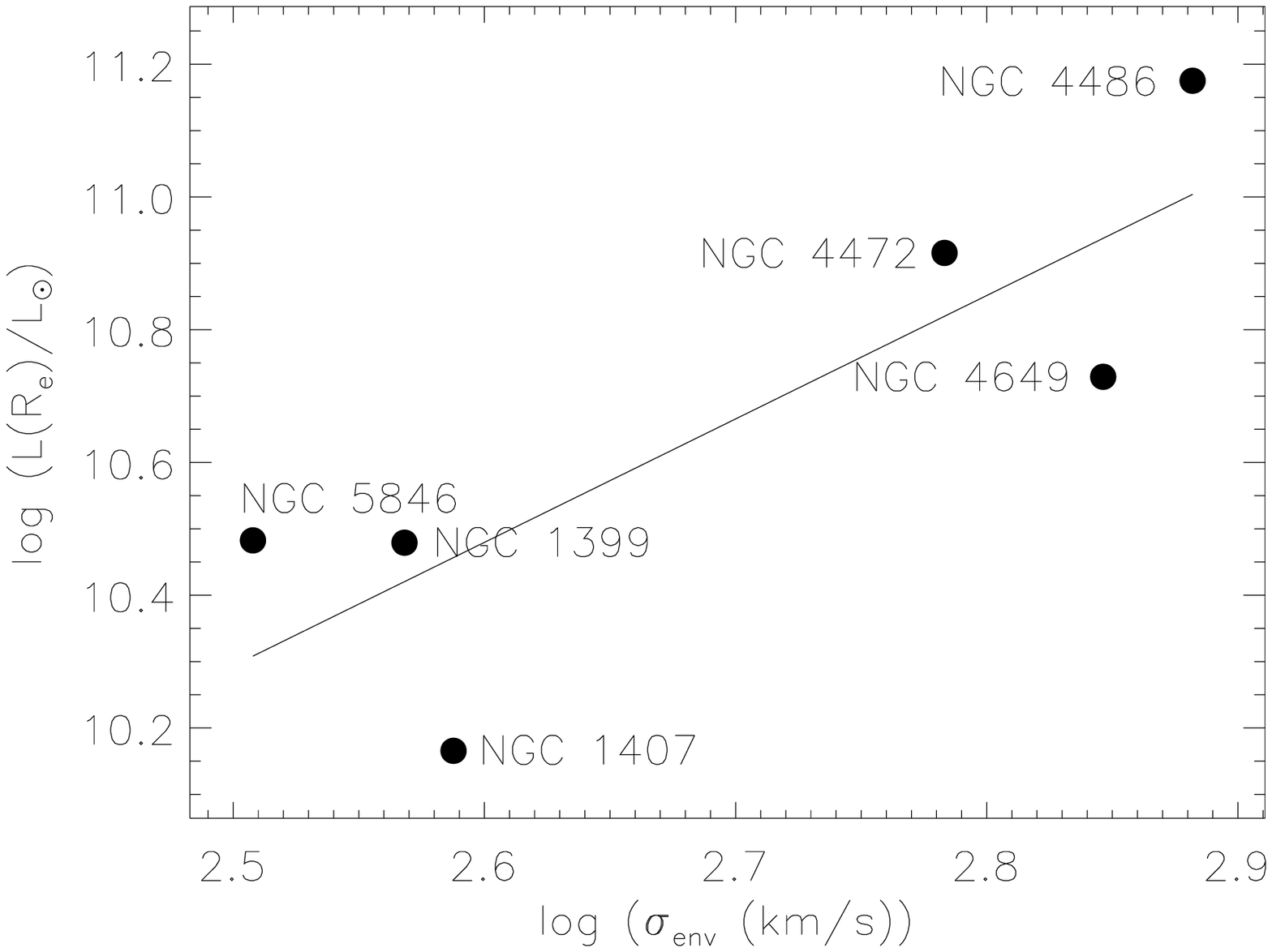}
}
\caption{Correlations between the stellar component, the total
  circular velocity, and the environment: (a) Tully-Fisher relation
  between the luminosity of the stellar component at $1R_e$ and the
  total circular velocity from the X-rays at $1R_e$ (black filled
  circles) and associated 95\% confidence range (horizontal
  bars). Also shown are the results of considering the average effect
  of the unresolved population of low-mass X-ray binaries and an
  abundance gradient (black open circles). (b) Total circular velocity
  from X-rays at $1R_e$ and the velocity dispersion of the environment
  (black filled circles). (c) Luminosity of the stellar component at
  $1R_e$ against the velocity dispersion of the environment (black
  filled circles). Solid black and dashed black lines show
  straight-line fits to the black filled circles and open circles
  respectively.  \label{fig:corr}}
\end{figure}

\subsection{Correlations between the total circular velocity curves,
  the stellar component and the environment}\label{sub:corr}

The Tully-Fisher relation links the luminous component to the total
mass of the galaxies and for spiral galaxies is usually shown between
the total luminosity and the flat part of the circular velocity
curve. We plot an analogous relation for our sample between the total
$B$-band luminosity at $R_e$ against the total circular velocity from
the X-rays at $R_e$, shown with black filled circles in Figure
\ref{fig:corr}. We have also plotted with black open circles, the
approximate total circular velocities at these radii if we take into
account the average effects of the unresolved population of low-mass
X-ray binaries and varying abundance profiles, as estimated by
\cite{chur+10}. The solid black and dashed black lines show straight
lines fit to the black filled circles and black open circles
respectively, considering the errors in the circular velocity. The
lines have slopes of $3.77 \pm 0.03$ and $4.01 \pm 0.03$, very similar
to the value of 4 in the Tully-Fisher relation for spiral galaxies and
to the value of $\sim 4$ ({\it private communication} with J. Thomas)
obtained for the sample of Coma cluster elliptical galaxies analysed
in \cite{thomas+09}. PMCC values for the fits are 0.86 and 0.85,
therefore supporting a linear correlation at a $\sim 95\%$ level of
significance for a sample size of six. This implies that systems with
more luminous stellar components tend to have higher circular
velocities.

Figure \ref{fig:corr}(b) shows the relation between the circular
velocity at $R_e$ and the velocity dispersion of the surrounding
environment from column (6) in Table \ref{tab:sample}. The solid line
shows the best-fit straight line with a slope of $0.53 \pm 0.02$ and a
PMCC value of 0.92, supporting a linear correlation at a $98 \%$ level
of significance. This strongly implies that central galaxies residing
in hotter environments (as measured by the velocity dispersions) have
higher circular velocities. Figure \ref{fig:corr}(c) shows the
logarithm of the luminosity at $R_e$ against the logarithm of the
velocity dispersion of the surrounding environment. The best-fit
straight line has a PMCC value of 0.83, supporting a linear
correlation at a level of $\sim 95\%$ signficance. This implies that
the luminosity of the galaxy residing at the centre of a sub-clump,
group or cluster environment, is related to the velocity dispersion of
the environment.

In environments where the local velocity dispersions are higher, the
densities ($\sim \sigma^2/r^2$) are also greater, as the radii (0.7--1
Mpc) over which these velocity dispersions were calculated are
comparable. Therefore one would expect that a larger number of systems
would fall onto the central galaxy, resulting in a more luminous
stellar component and a more massive dark matter
halo. \cite{brough+07} found that more X-ray luminous systems are
intrinsically brighter in the $K$-band compared to less luminous X-ray
luminous systems for their sample of three brightest group galaxies
and three brightest cluster galaxies. In denser environments, one
would expect the gas to be hotter and denser, and therefore more X-ray
luminous, thereby corroborating the results of \cite{brough+07}.

\section{CONCLUSIONS}\label{sec:conc}

In this paper, we describe a new non-parametric Bayesian approach to
obtain mass distributions and associated confidence ranges from
temperature and density profiles of hot gas in hydrostatic
equilibrium. The method is able to successfully reconstruct the mass
distribution of a test galaxy just within a 68\% confidence range of
the recovered models. We assume a smoothing prior to ensure unique and
physical mass distributions and calibrate this on the test galaxy.

We apply the procedure to the sample of six X-ray bright elliptical
galaxies from \cite{chur+10}, who used high-quality X-ray observations
of Chandra and XMM-Newton to obtain temperature and density profiles
of the hot gas. We find total mass distributions with an average mass
of $\sim 1.5\times 10^{12} M_{\odot}$ and average circular velocity of
$\sim 515$ km/s at a radius of 25 kpc.

The total circular velocity curves of our sample are all rising in the
outer parts as a result of both an increasing temperature profile and
a logarithmic pressure gradient that generally increases in magnitude
outwards. Therefore the mass distributions of our sample are not
isothermal, suggesting that we are probing the mass distribution
sufficiently far out to observe the effects of the massive
group/cluster haloes in which these galaxies reside. Increasing
circular velocity curves appear to contradict work done in the past
but are most likely a consequence of: i) Samples in the literature not
restricted to only massive elliptical galaxies and/or ii) mass
profiles not being probed sufficiently far out.

The stellar components of our sample show a remarkable coherence in
their surface-brightness profiles when scaled by the radius where
$d\mu /d\log R = 2$. We find however that the effective radius is a
better proxy for the luminosity of the stellar component. The dark
matter mass fraction is 35--$80\%$ at $2R_e$ and increases to about
80--$90\%$ at the outermost radii.

Except in the case of NGC 1407, our determination of circular velocity
curves from X-rays agrees well with previous X-ray
determinations. When comparing our circular velocity curves with those
found by dynamical models in the literature, we find that the X-ray
circular velocity curves are often lower, particularly in the central
10 kpc. This is probably due to contributions from: i) Non-thermal
pressures, ii) multiple-temperature components, iii) incomplete
spatial coverage in the kinematic data used for the dynamical models,
and iv) insufficiently general mass profiles explored in the dynamical
modelling.

Looking at relations between the luminosity of the stellar component
at $R_e$, the total circular velocity at $R_e$ and the velocity
dispersion of the environment, we find evidence for a dependence of
the properties of central galaxies on their environment.

The next steps in this project are to extend the analysis to a larger
sample of X-ray bright elliptical galaxies to examine the
isothermality of their circular velocity curves, and to obtain more
stringent constraints on how they compare with alternative methods of
mass determinations, in particular through the construction of more
radially extended dynamical models. This will enable us to build up a
clearer picture of the connection between properties of central X-ray
bright elliptical galaxies and the environment in which they reside.

\section*{ACKNOWLEDGEMENTS}\label{sec:ack}

PD was supported by the DFG Cluster of Excellence ``Origin and
Structure of the Universe''. EC and IZ were supported by the DFG grant
CH389/3-2 and OFN-17 program of the Russian Academy of Sciences. PD
and IZ would like to thank the International Max Planck Research
School (IMPRS) in Garching.

\bibliographystyle{mn2e}
\bibliography{paper}

\appendix

\label{lastpage}

\end{document}